\documentstyle[twoside,fleqn,jhuwkshp,psfig]{article}
\newcommand{\etal}{{\it et al.} }
\newcommand{\asca}{{\it ASCA} }

\newcommand{\xmm}{{\it XMM-Newton} }
\newcommand{\chandra}{{\it Chandra} }

\newcommand{\astroe}{{\it Astro-E2} }
\newcommand{\rxte}{{\it RXTE} }
\newcommand{\hetg}{{\it HETGS} }
\newcommand{\rrg}{$r/r_{g}$ }

\newcommand{\fekalfa}{{Fe~K$\alpha$} }

\newcommand{\fexxv}{Fe~{\sc xxv} }
\newcommand{\fexxvp}{Fe~{\sc xxv}}
\newcommand{\fexxvi}{Fe~{\sc xxvi} }
\newcommand{\fexxvip}{Fe~{\sc xxvi}}
\newcommand{\feklya}{{Fe~{\sc xxvi }~Ly$\alpha$} }
\newcommand{\feklyb}{{Fe~{\sc xxvi}~Ly$\beta$} }
\newcommand{\feklybp}{{Fe~{\sc xxvi}~Ly$\beta$}}
\newcommand{\feklyap}{{Fe~{\sc xxvi}~Ly$\alpha$}}
\newcommand{\bsax}{{\it BeppoSAX} }
\newcommand{\figlcurves}{{Fig.~1} }
\newcommand{\figlcurvesp}{{Fig.~1}}

\newcommand{\fighegfek}{{Fig.~2} }
\newcommand{\fighegfekp}{{Fig.~2}}
\newcommand{\figcontours}{{Fig. 3} }
\newcommand{\figcontoursp}{{Fig. 3}}
\newcommand{\fighorns}{{Fig.~4} }
\newcommand{\fighornsp}{{Fig.~4}}

\newcommand{\fighornse}{{Fig.~4(e)} }

\newcommand{\figwidthshift}{{Fig.~5} }
\newcommand{\figwidthshiftp}{{Fig.~5}}

\newcommand{\figrxtep}{{Fig.~6}}
\newcommand{\figrxtea}{{Fig.~6(a)} }
\newcommand{\figrxteap}{{Fig.~6(a)}}
\newcommand{\figrxteb}{{Fig.~6(b)} }

\newcommand{\figrxtec}{{Fig.~6(c)} }

\newcommand{\tablehegfits}{Table~1 }
\newcommand{\tablehegfitsp}{Table~1}
\newcommand{\tablepcafits}{Table~2 }
\newcommand{\tablepcafitsp}{Table~2}

\newcommand{\src}{NGC~7314 }

\begin{document}

\title{\bf Fe~{\sc xxv} and Fe~{\sc xxvi} Diagnostics of the Black Hole
and Accretion Disk in Active Galaxies: Chandra Time-Resolved
Spectroscopy of NGC~7314}

\author{Tahir Yaqoob\address{\it Department of Physics and Astronomy,
Johns Hopkins University, Baltimore, MD 21218.}
\address{\it Laboratory for High Energy Astrophysics,
NASA/Goddard Space Flight Center, Greenbelt, MD 20771.},
Ian M. George$^{\rm b \ }$\address{\it Joint Center for Astrophysics, University of Maryland,
Baltimore County, 1000 Hilltop Circle, Baltimore, MD21250.},
Timothy R. Kallman$^{\rm b}$,
Urmila Padmanabhan$^{\rm a}$,
Kimberly A. Weaver$^{\rm b}$,
T. Jane Turner$^{\rm b \ c}$}

\begin{abstract}
\vspace{-6mm}
\centerline{\bf Abstract}

We report the detection of \fexxv and \fexxvi $K\alpha$ emission lines
from a \chandra High Energy Grating Spectrometer (\hetg)
observation of the narrow-line Seyfert~1 galaxy NGC~7314,
made simultaneously with {\it RXTE}.
The lines are redshifted ($cz \sim 1500 \rm \ km \ s^{-1}$)
relative to the systemic velocity
and unresolved by the gratings. 
We argue that the lines originate in a near face-on ($<7^{\circ}$)
disk having a radial line emissivity flatter than
$r^{-2}$. Line emission from ionization states of Fe
in the range
$\sim$~Fe~{\sc i} a
up to \fexxvi is observed. The ionization balance of Fe responds
to continuum variations on timescales less than 12.5~ks,
supporting an origin of the lines 
close to the X-ray source. We present additional, detailed
diagnostics from this rich data set.
These results identify NGC~7314 as a key source to
study in the future if we are
to pursue reverberation mapping of space-time near black-hole
event horizons. This is because
it is first necessary to
understand the ionization
structure of accretion disks and the relation between
the X-ray continuum and \fekalfa line emission.
However, we also describe how our results are suggestive of
a means of measuring black-hole spin without a knowledge
of the relation between the continuum and line emission.
Finally, these data emphasize that
one {\it can} study strong gravity with narrow
(as opposed to very broad) disk lines.
In fact narrow lines offer higher precision, given sufficient
energy resolution. 

 
\begin{center}
{\it Accepted for Publication in the Astrophysical Journal 20 June 2003. Submitted 22 Jan 2003}
\end{center}

\end{abstract}
\maketitle

\pagebreak

\section{INTRODUCTION}
\label{intro}

At least part of the 
\fekalfa fluorescent 
emission line in active galactic nuclei (AGN)
is believed to originate in a
relativistic accretion disk around a black hole
(e.g. see reviews by Fabian \etal 2000; Reynolds \& Nowak 2003).
In order to ultimately use the \fekalfa lines to
probe the space-time around a black hole, it has
been recognized that it will be necessary to understand
the complex ionization physics of the accretion disk
(e.g. Nayakshin \& Kallman 2001; Ballantyne, Ross, \& Fabian 2001;
Ballantyne \& Ross 2002). In
particular, we must understand 
the relation between the continuum and line emission.
The basic observational data to build such an understanding
is sparse. One needs to identify sources with rapidly variable
line and continuum emission, in which the dynamic range
of the continuum variability is large. 
Once the emission-line variability is firmly
established, the ground-work can be prepared for future missions
to use these sources for time-resolved spectroscopy to probe shorter
and shorter timescales,  with higher spectral resolution.
 
Even in the first study of the \fekalfa line emission in 
a sample of type~1 AGN (Nandra \etal 1997), the variety
in the shape and width of the \fekalfa  line was pointed out.
The range in width has been confirmed by subsequent studies
(e.g. Lubi\'{n}ski and Zdziarski 2001; Yaqoob \etal 2002;
Perola \etal 2002; Reeves 2002; Yaqoob \& Padmanabhan 2002).
We have now observed \fekalfa lines with
FWHM less than 40~eV, to more than a couple of keV.
It has been traditional to associate `narrow' \fekalfa lines   
with an origin in distant matter, at least several
thousand gravitational radii from the black hole (e.g.
the optical broad-line region (BLR), the
putative obscuring torus, or the optical narrow-line region (NLR)).
However, Petrucci \etal (2002) recently reported a {\it variable},
narrow \fekalfa line in Mkn~841, supporting an accretion-disk origin.
In addition, it has been pointed out that \fekalfa lines
which appear to be narrow even with
the highest spectral resolution instruments
currently available (the \chandra gratings), may have
a significant contribution from the accretion disk
(Lee \etal 2002; Yaqoob \& Padmanabhan 2002).
While such lines may be interpreted in terms of
a truncated disk (e.g. Done, Madejski, \& \.{Z}ycki 2000), they could be
due to a flat radial line emissivity (i.e. 
intensity per unit area falling off with radius more
slowly than $r^{-2}$). 

In this paper we report on the results of a 
high spectral resolution \chandra grating 
observation of the narrow emission-line Seyfert~1 galaxy 
(NELG) NGC~7314 in
which we detect \fekalfa emission from
not only the low ionization states of Fe (Fe~{\sc xvii}
or lower), but also from
redshifted He-like, and H-like Fe.
We find that
the ionization balance of Fe responds to rapid continuum variations,
ruling out a distant-matter origin for the high ionization-state lines.
We are able to use these line diagnostics to infer 
quantitative, new information about the black-hole/accretion disk
system in NGC~7314, opening up similar possibilities
for other AGN. 

NGC~7314 is known for its rapid, large amplitude 
X-ray variability (Turner 1987; Yaqoob \etal 1996; 
Branduardi-Raymont \etal 2002). 
The luminosity can change by a factor
of $\sim 2$ on a timescale of hundreds of seconds and
by a factor of $\sim 4$ in thousands of seconds. 
The luminosity itself is fairly low: even at the
peaks of the largest flares, the 2--10 keV luminosity
is only $\sim 2 \times 10^{42} \rm \ ergs \ s^{-1}$
(we assume $H_{0} = 70 \rm \ km \ s^{-1} \ Mpc^{-1}$ and $q_{0}=0$ throughout). 
It is well known that the least luminous
Seyfert~1 galaxies are the most variable 
(e.g. Green, McHardy, \& Lehto 1993; Turner \etal 1999)
and measurements of the mass of the putative central
black hole yield a correspondingly small value
of $\sim 5\times 10^{6}M_{\odot}$ (Padovani \& Rafanelli 1988;
Schulz, Knake, \& Schmidt-Kaler 1994).
NGC~7314 was also one of the first AGN in which rapid
variability of the Fe-K emission complex was found (Yaqoob \etal 1996).
However, the limited spectral resolution of \asca
restricted the 
diagnostics that could be made using this variability.
The main result was that the emission in the red and blue wings
of the line profile responded to the continuum intensity,
whilst the core emission at 6.4~keV did not respond.
Thus, we observed NGC~7314
with the highest spectral resolution available in the Fe-K band,
using the \chandra gratings, in order elucidate the nature of
the line emission and its variability. The observation was
made simultaneously with \rxte in order to measure the hard
X-ray continuum as well, since it can provide additional constraints
on the accretion disk.

In this paper we adopt $z=0.004760$ for
the redshift of NGC~7314, which was obtained from 
H~{\sc i} measurements (Mathewson \& Ford 1996).
We shall see that the \chandra data are sensitive to
uncertainties in $z$ of $0.00022$, but this is
larger than the uncertainty quoted by
Mathewson \& Ford (1996), of $\pm 0.000033$.
The Galactic column density along the line-of-sight to NGC~7314 is
small, and we use the value $1.46 \times 10^{20} \ \rm cm^{-2}$
(Dickey \& Lockman, 1990) throughout this paper. 

The paper is organized as follows.
In \S\ref{data} we describe the observations, data, and
methods of analysis. 
In \S\ref{hegspec} we present the results of 
high spectral resolution, time-resolved
spectroscopy of the Fe-K region, revealing
a wide and variable mix of ionization states of Fe. 
In \S\ref{releffects} we 
describe how the measurement of
relativistic effects from the X-ray emission-lines
can be used to constrain the structure of the 
accretion disk.
Then in \S\ref{pcafits} we discuss how
simultaneous \rxte data add further constraints on the
accretion disk. In \S\ref{missions} we discuss 
the implications of our results for AGN in general,
and how future instrumentation could be used to advance our
knowledge further.
Finally, in \S\ref{summary} we
summarize our results.

\section{OBSERVATIONS AND DATA}
\label{data}
 
We observed \src with \chandra in July 2002, simultaneously
with {\it RXTE}. The \chandra observation was made in two
parts, on July 19, from UT 04:26:00 to UT 12:45:44, and 
on July 20, from UT 03:18:16 to UT 22:42:19. The \rxte observation
was made from July 19 UT 04:39:21 to July 22 UT 16:26:34.

For this study the \chandra observation was
made with the High-Energy Transmission Grating (or \hetg --
Markert, \etal 1995) in the focal plane of the
High Resolution Mirror Assembly. The \chandra \hetg affords 
the best spectral resolution in the $\sim 6-7$~keV Fe-K band currently
available ($\sim 39$~eV, or $1860 \rm \ km \ s^{-1}$ FWHM
at 6.4 keV). \hetg consists of two grating assemblies,
a High-Energy Grating (HEG) and a Medium-Energy Grating (MEG),
and it is the HEG which achieves this spectral resolution.
The MEG spectral resolution
is only half that of the HEG. The HEG also has
higher effective area in the Fe-K band, so our study will
focus principally on the HEG data.
The HEG and MEG energy bands are 0.4--10~keV and 0.7--10~keV
respectively, but the effective area falls off rapidly with energy
near both ends of each bandpass. 
The \chandra data 
were processed with version 6.8.0
of the processing software, {\tt CALDB} version 2.15
was used, and the telescope responses made using
{\tt ciao 2.1.3}\footnote{http://asc.harvard.edu/ciao2.1/documents\_threads.html}. 
Otherwise,
HEG and MEG lightcurves and spectra
were made exactly as described in Yaqoob \etal (2003). 
We used only the first orders of the grating data (combining
the positive and negative arms). The zeroth order data
were piled up and the higher orders contain much fewer counts
than the first order. The mean count rates
(in the full energy band of each grating), over the entire
\chandra observation were $0.1965 \pm 0.0014$ ct/s and
$0.3532 \pm 0.0019$ ct/s for HEG and MEG respectively. 
HEG and MEG spectra extracted over the entire observation,
resulted in a net exposure time of 97.248 ks.
Background was not subtracted since it is negligible above
$\sim 0.6$~keV (see also \S\ref{hegspec} and \fighegfekp [a]).

The \rxte PCA data
were reduced using methods described in 
Weaver, Krolik, \& Pier (1998), except that we used a
later version of the spectral response matrix generator (V 8.0) and
the latest version of 
background model (`L7\_240\_FAINT') released in
February 2002, with {\tt ftools v5.2}. Only data from layer 1 were
used. Of the five PCUs of the PCA, fewer were operated later 
in  the mission than earlier, and we obtained useful
data only from PCU0 and PCU2.
We extracted background-subtracted lightcurves and spectra 
from each PCU separately.
For weak sources, such as AGN, the PCA
spectra may be unreliable above $\sim 15$ keV, when the
background-subtraction systematics become too large. 
The usable low-energy limit of the spectra is
$\sim 3$~keV, below which the instrument response is
too uncertain.
For spectra extracted over the entire \rxte observation
(which continued beyond the \chandra observation),
the exposure time was 82.128~ks (each PCU) and
the mean 3--15~keV background-subtracted count rates
were $3.415 \pm 0.015$ ct/s and $2.994 \pm 0.011$ ct/s
per PCU, for PCU0 and PCU2 respectively. However, in this paper
we will use only the \rxte data which overlapped with
\chandra (see \S\ref{intselect}).

\subsection{\chandra and \rxte Cross-Calibration}
\label{crosscal}

Model-fitting cannot be performed on contemporaneous 
\chandra and \rxte data simply by spectral fitting
to the spectra from different instruments simultaneously.
This is because the effective area of the \rxte PCA is so much
larger than the \chandra \hetg that the former will
bias the fits (PCA:HEG ratio is $>40$ per PCU at 6.4~keV). 
Also, the spectral resolution of the PCA
is more than a factor of 30 worse than the HEG in the Fe-K band.
Therefore we will take the approach of spectral fitting data from the
different instruments independently
and use any 
line emission measurements from the \chandra \hetg as
fixed components of any model used to perform spectral
fitting on the PCA data. In the case of complex line
emission, all the parameters measured from the \chandra
data have to be fixed, otherwise the PCA spectral fits
become unstable, as the PCA can confuse line emission
and continuum due to the poor spectral resolution.
In order to execute this approach we must know at least the
cross-normalization between the \chandra and \rxte instruments.   
This is because all line intensities measured by \chandra
have to be adjusted before they can be applied to the PCA data.

Although cross-calibration studies have shown that
\chandra fluxes generally agree well with
other missions\footnote{http://space.mit.edu/ASC/calib/hetgcal.html},
it is known that in the 3--10 keV band
PCA fluxes are systematically higher than 
\asca and \bsax fluxes
\footnote{ASCA GOF calibration memo at \\
http://heasarc.gsfc.nasa.gov/docs/asca/calibration/3c273\_results.html}.
However, these PCA cross-calibration
results are out of date since new background-subtraction and
calibration for PCA data have since been introduced.
Therefore we studied a simultaneous \chandra/\rxte
observation of the
Seyfert~1 galaxy Fairall~9, observed in 2001, September~11.
These data sets are available from
the public archives and were selected
because in these observations the source had an X-ray spectrum which was
a simple power law in the 0.6--15~keV band
(after accounting for Galactic absorption), no 
spectral features apart from an \fekalfa line in the
6--7 keV band, and 
no X-ray variability was detected. We measured
the excess variance (e.g. see Turner \etal 1999) from the \chandra \hetg data
(0.8--7~keV for HEG, 0.6--5~keV for MEG, binned at 128~s)
to be $(1.5 \pm 6.5) \times 10^{-4}$.

PCU0 did not yield any useful data for the
Fairall~9 observation so we used only PCU2 for
the cross-calibration of flux, extracting a spectrum
over the entire observation (yielding an exposure time of
66.928~ks). Since we will only use
the HEG data for studying the Fe-K region in NGC~7314, we
extracted a HEG first-order spectrum over the entire
observation of Fairall~9 (yielding an exposure time of 78.893~ks).
We fitted the 3--9 keV HEG and PCA spectra independently with a simple
power-law model plus Gaussian emission line and
obtained good fits for both. The energy range 
used is accessible to both instruments. The HEG spectrum 
gave a photon index which was steeper by $\sim 0.1$ than
that obtained from the PCA spectrum. 
We then measured the 6--7 keV fluxes from 
the best-fitting models and found that
line intensities measured by the \chandra HEG 
must be increased by 14.7\% if they are to be applied
to PCA PCU2 data.
Also, we compared the normalization of PCU0 with PCU2
using our NGC~7314 data and found a 3\% difference
in the normalization between the two PCUs.
Therefore, if we are using combined PCU0 and PCU2 spectra
with equal weighting, \chandra HEG line intensities
must be increased by 13.0\% and we adopted this
latter value for cross-normalization of emission lines
in the Fe-K region. Note that there have been no
changes in the calibration of either the HEG or PCA
over the period between the Fairall~9 and NGC~7314
observations that would significantly affect this result.

\subsection{Lightcurves and Intensity-Selected Spectra}
\label{intselect}

\begin{figure*}[tbh]
\vspace{10pt}
\centerline{\psfig{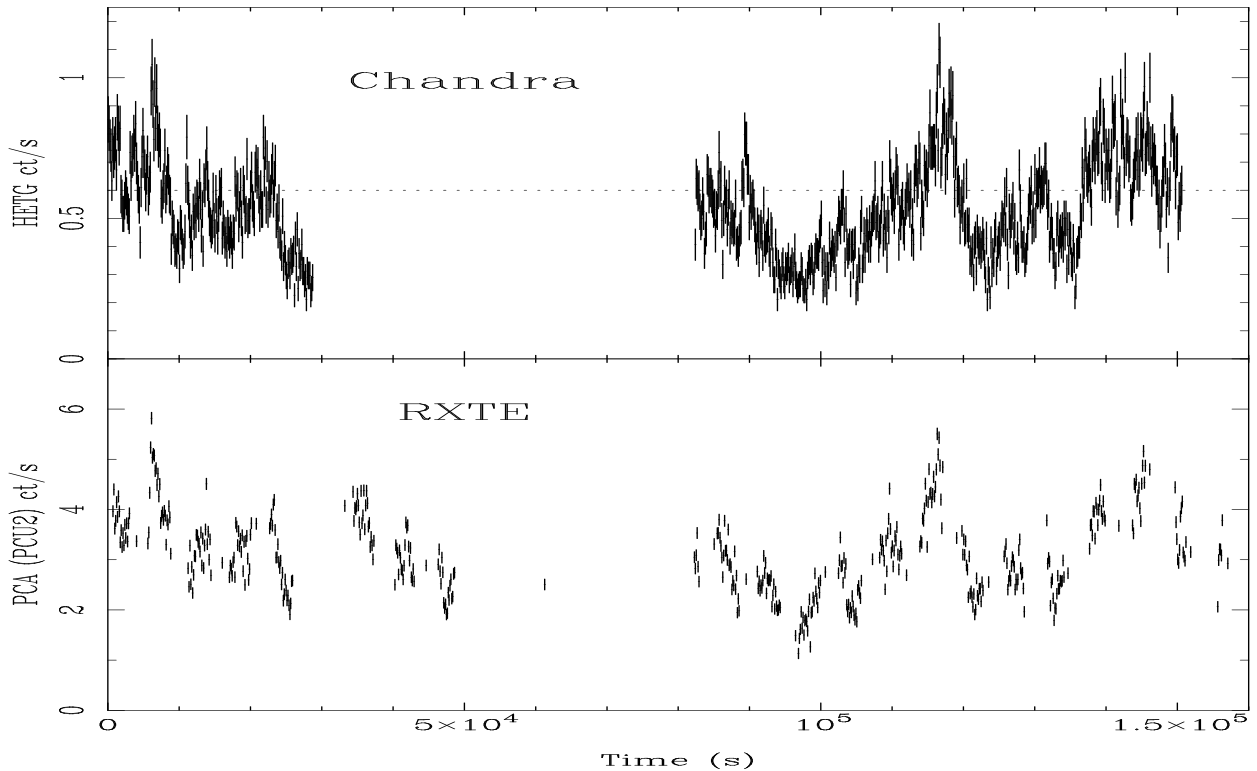}}
\caption{
\chandra \hetg and \rxte PCA lightcurves of NGC~7314
binned at 128s.
The \hetg data  are from combined HEG and MEG
$\pm 1$ orders (0.8--7~keV).
The PCA data are from PCU2 only (3--13~keV).
The reference time corresponding to  $t=0$ for both
lightcurves is 2002 July 19, UT 04:33:36.}
\end{figure*}

\begin{figure*}[tbh]
\vspace{10pt}
\centerline{\psfig{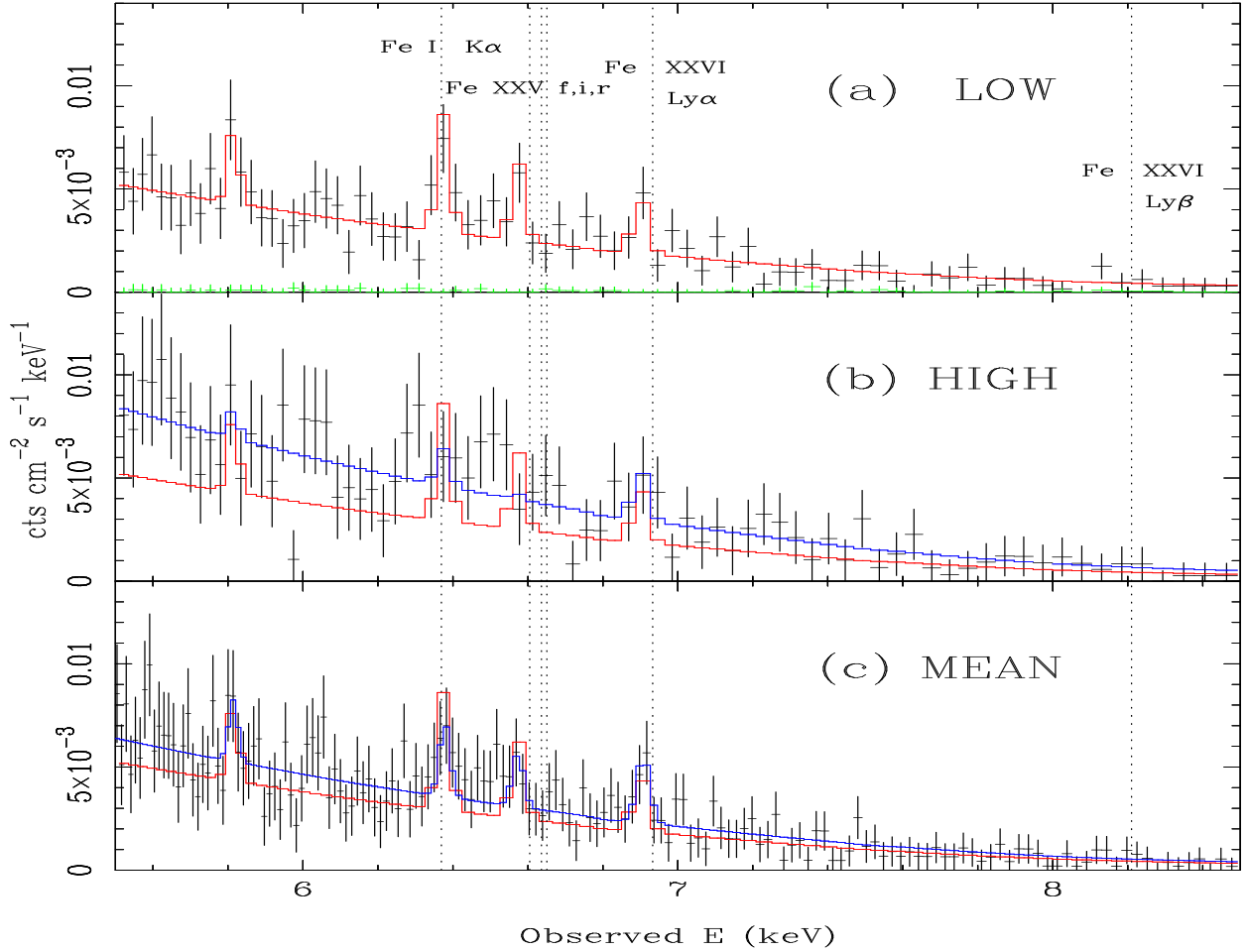}
}
\vspace{-1cm}
\caption{
\chandra HEG low-state, high-state, and mean spectra in the 5.5--8.5~keV
band. The bin size for the low-state and high-state spectra is
$0.01\AA$ but $0.005\AA$ for the mean spectrum (the HEG
FWHM resolution is $0.012\AA$). The spectra have not
been corrected for instrument response so that no
artifically `sharp' features are introduced into the plot:
instead the models have been folded through the instrument
response. The red solid lines in each
panel corresponding to the best-fitting power-law plus four-Gaussian
model to the {\it low-state} spectrum. The blue solid lines
are the best-fitting models for the data shown in the appropriate panel
(i.e. high-state and mean spectra).
The normalized background spectrum extracted from a region 3.6--10 arcsec
either side of the dispersion axis is shown in green in panel (a).
In the low state we detect emission lines from
Fe~{\sc i}--Fe~{\sc xvii}~$K\alpha$, \fexxv (f), \feklyap, and
\feklybp. The dotted lines show expected energies in the observer's frame
and it can be seen that
the latter three lines are redshifted relative to the
systemic velocity (see \S\ref{hegspec}).
None of the four emission lines modeled are resolved.
In the high state the continuum has increased
by a factor $\sim 1.7$ compared to the low state.
In the high state, additional \fekalfa emission appears redward of
the energy of Fe~{\sc i}~$K\alpha$,
as well as
some unresolved \fekalfa emission
from ionization intermediate between Fe~{\sc i}~$K\alpha$ and \fexxvp.
This response to
the continuum is rapid ($<12.5$~ks). See also \figcontoursp.
The emission line at 5.84~keV
(rest-frame) remains unidentified, but could be due relativistically
shifted \fekalfa emission (see \S\ref{lowefeature}).
}
\end{figure*}

\begin{figure*}[tbh]
\vspace{-1cm}
\vspace{10pt}
\centerline{\psfig{file=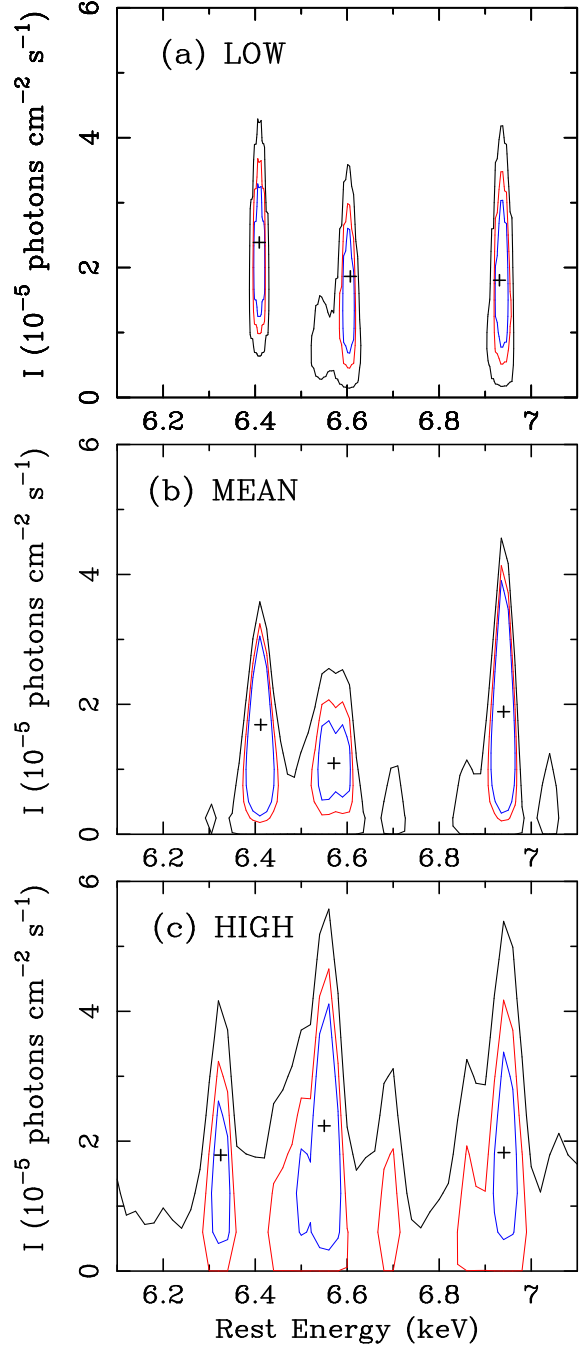,angle=180.0,width=6.5in,height=8.0in}
}
\vspace{-1.75cm}
\caption{
Confidence contours of line intensity versus
energy, showing the rapid ($<12.5$~ks) variability of the
Fe-K emission complex in NGC~7314 (see also \fighegfek). The two-parameter joint
confidence levels correspond
to 68\% (blue), 90\% (red), and 99\% (black). In the low state there are
three clear, discrete emission features corresponding to $K\alpha$
line emission from Fe~{\sc i}--Fe~{\sc xvii}, \fexxvp, and \fexxvip.
Note that the signal-to-noise of the low-state and high-state
spectra is approximately the same and the signal-to-noise of the
mean spectrum is approximately twice that of either low-state or high-state
spectra alone. Therefore the differences between the
contours in different states are {\it not} due
to signal-to-noise effects. This is supported by the fact that
the centroid and width of the emission near 6.6~keV
varies significantly (this is clearest from a comparison of
the low-state and mean spectra).
As the continuum level increases, the
ionization structure of the emitting region changes from
producing three discrete \fekalfa features to a more complex
blend of emission.
In the high state, it also appears that the
\feklya emission becomes broader,
and even the lowest ionization emission component
appears to be variable: a peak below
6.4~keV becomes more  prominent than the peak at $\sim 6.4$~keV.
}
\end{figure*}

\begin{figure*}[tbh]
\vspace{10pt}
\centerline{\psfig{file=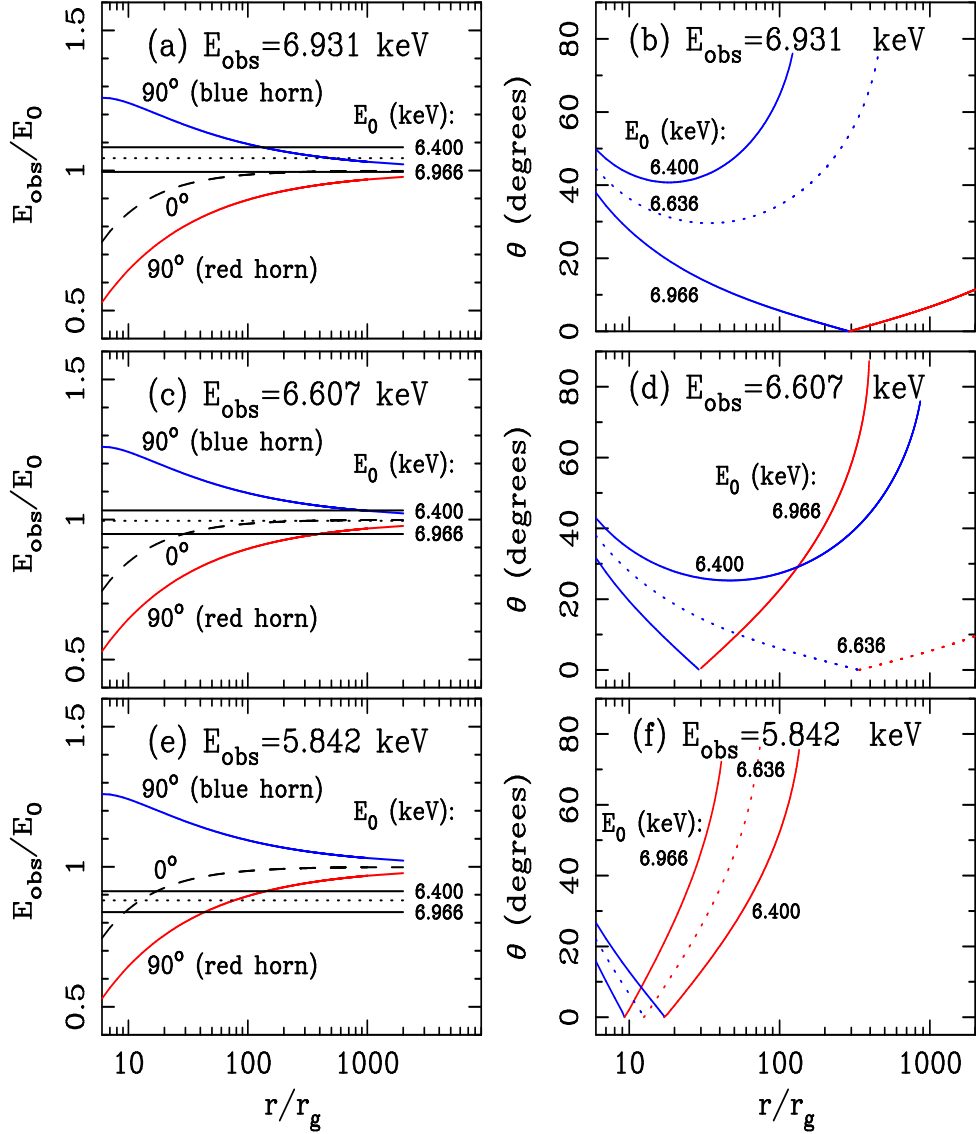,angle=270.0,width=6.5in,height=6.0in}
}
\vspace{-1cm}
\caption{
Constraints on the disk inclination angle ($\theta$) and the
radius (in units of $r_{g} \equiv GM/c^{2}$) from \fekalfa emission
lines originating from hot spots or annuli on a Keplerian disk
rotating around a central Schwarzschild black hole
(see \S\ref{narroworigin}).
Pairs of panels are shown for each of the three strongest
narrow lines observed in the low-state spectrum of NGC~7314,
at observed energies ($E_{\rm obs}$) of 6.931, 6.607, and 5.842~keV.
Horizontal lines in panels (a), (c), and (e), and the curves in
(b), (d), and (f) are labeled according to values of the
line rest energy, $E_{0}$, of 6.400 (Fe~{\sc i}~$K\alpha$),
6.636 (\fexxv (f): dotted), and 6.966~keV (\feklyap).
Blue curves indicate approaching parts of the disk and
red curves indicate receding parts.
For a given disk inclination angle, there are a pair of
curves (blue and red) of $E_{\rm obs}$/$E_{0}$
versus $r/r_{g}$, lying inside the $90^{\circ}$ envelopes
in panels (a), (c), and (e). The observed value of
$E_{\rm obs}$/$E_{0}$ would then vary (as a hot spot
rotates with the disk) along a vertical
line placed at a value of $r/r_{g}$ corresponding to its
location on the disk, between the appropriate blue and red curves.
An annulus of emission, in which the observed
lines are identified with either the red or blue Doppler horns,
occupies a unique point in panels (a), (c), and (e), with
specific values of $\theta$, $r/r_{g}$, and $E_{\rm obs}/E_{0}$
(if two of these quantities are known, there is no choice for
the third).
If the observed emission lines do indeed correspond
to red or blue Doppler horns, panels (b), (d), and (f)
show the necessary relations between $\theta$ and $r/r_{g}$ for three
values of $E_{0}$.
}
\end{figure*}

\begin{figure*}[tbh]
\vspace{-1cm}
\vspace{10pt}
\centerline{\psfig{file=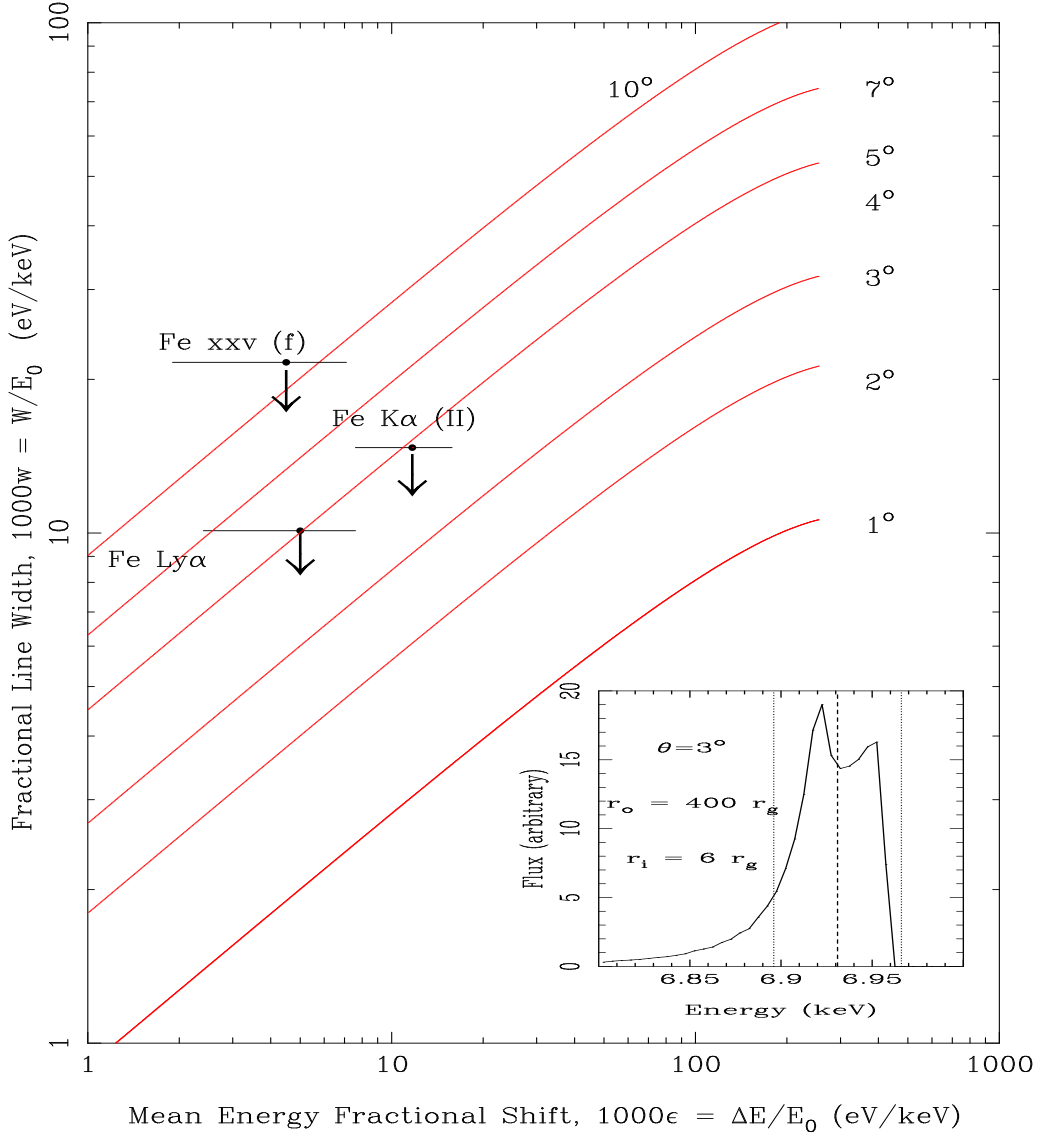,width=6.5in,height=7.0in}
}
\vspace{-1.25cm}
\caption{
The relation between the minimum width ($w \equiv W_{\rm min}/E_{0}$),
and energy shift
($\epsilon \equiv \Delta E/E_{0}$) of a monochromatic emission line
(rest-energy, $E_{0}$), originating in
a Keplerian disk rotating around a Schwarzschild black hole,
for the case that the red and blue Doppler horns are
unresolved (see \S\ref{widthshift}).
Plotted is $1000w$ versus
$1000\epsilon$, for several inclination angles (solid lines).
Note that, due to gravitational redshifting,
the line centroid can never be blueshifted relative to the rest-energy.
Each curve goes from $10^{4}$ to 6 gravitational
radii, from left to right, respectively.
An emission line resulting from integration between {\it any two radii}
on a disk with a given inclination angle must
have measured widths and energy shifts (relative to the rest-energy) which lie
{\it above} the curve for that inclination angle.
The measurement of the upper limit on the width of \feklya alone
(it is unresolved)
is firm enough to immediately rule out disks with $\theta >7^{\circ}$
since the measurements do not lie above curves with $\theta >7^{\circ}$.
The \fexxv (f) line
is also consistent with $\theta <7^{\circ}$. The implied maximum radius
of emission
(inferred from the \feklya line) is $\sim 600r_{g}$.
Also shown (inset) is
the line profile expected if the \feklya line were emitted uniformly
(constant intensity per unit area)
between 6 and 400 gravitational radii from a disk inclined at
$3^{\circ}$. The dashed line shows
the line centroid (as estimated simply from the mean energy of the
two Doppler peaks), and the dotted line shows the FWHM upper
limit measured from the low-state spectrum (see \tablehegfitsp).
The energy of the monochromatic model line is 6.966~keV in
the disk frame, and the observed centroid of the
relativistic line profile agrees with the measured centroid
energy of 6.93~keV in the low state (see \tablehegfitsp).
}
\end{figure*}

\begin{figure*}[tbh]
\vspace{10pt}
\centerline{\psfig{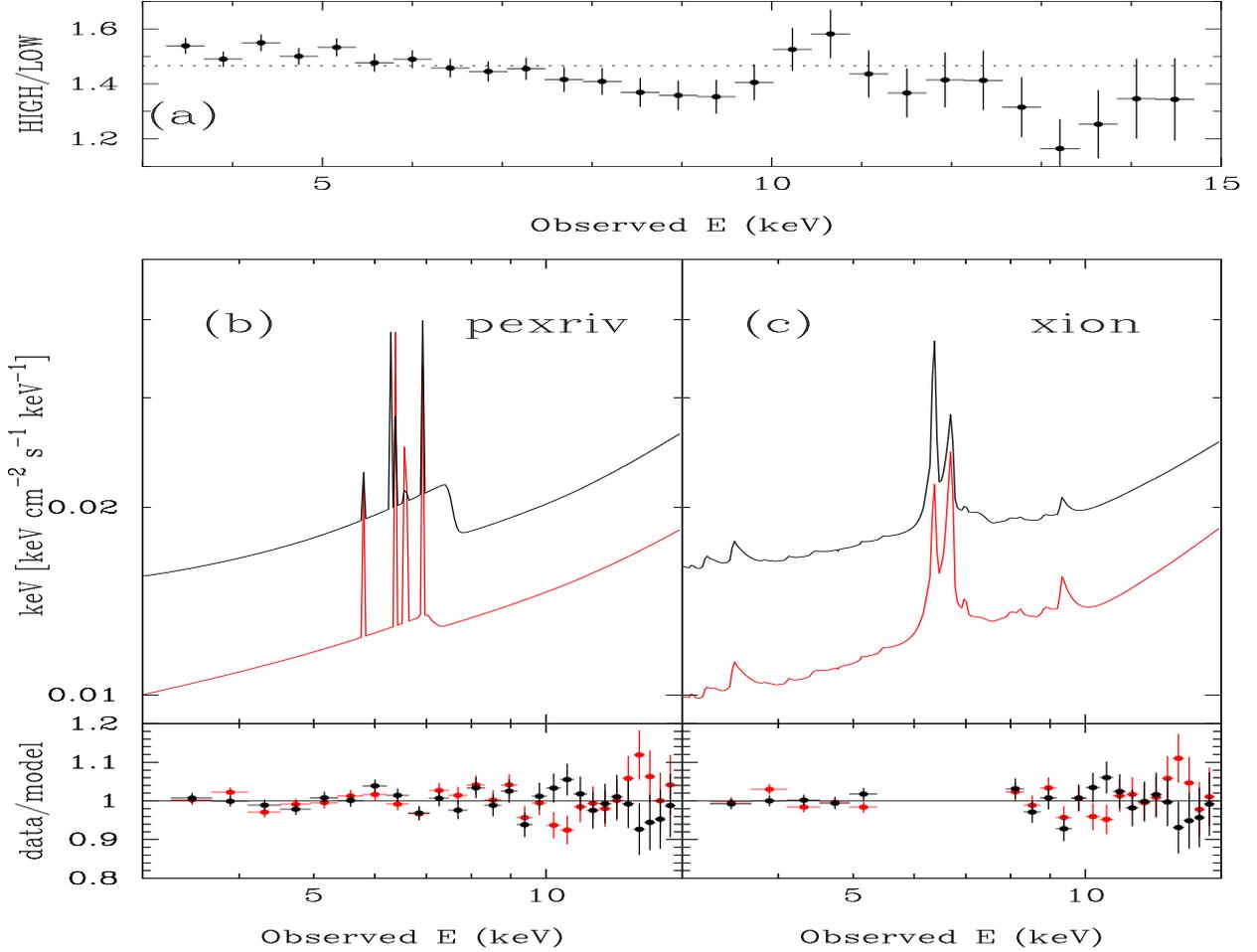}
}
\caption{
Results of time-resolved spectroscopy using the \rxte
PCA data (combining PCU0 and PCU2). Details in \S\ref{pcafits}.
(a) Ratio of 3---15~keV high-state spectrum to low-state
background-subtracted PCA spectra. The dotted line corresponds
to the ratio of high-state to low-state 3--15 keV background-subtracted count rates.
(b) Best-fitting ionized disk reflection models to the low-state (red)
and high-state (black) PCA spectra, using the XSPEC model {\sc pexriv}
(Magdziarz \& Zdziarski 1995). The Gaussian emission-line components
were frozen at the best-fitting \chandra values (\tablehegfitsp),
corrected for
the systematic difference in \chandra HEG and
\rxte PCA normalizations (see \S\ref{crosscal}).
The lower panel shows the corresponding data/model ratios.
(c) Best-fitting ionized disk reflection models to the low-state (red)
and high-state (black) PCA spectra, using the XSPEC model {\sc xion}
(Nayakshin \& Kallman 2001). The 5.5--7.5~keV data were omitted in the
fits since the model does not allow disk inclination angles
$<18^{\circ}$ and does not include the effects of turbulence
so predicts emission from H-like Fe to be much weaker than
that from He-like Fe (we observe similar equivalent widths for both --
see \fighegfek and \tablehegfitsp). The best-fitting models illustrate the line emission
predicted by {\sc xion} (relativistic smearing was turned off for
the plot in order to show clearly the relative strengths of the
lines from different ionization states).
}
\end{figure*}

\figlcurves shows the \chandra \hetg
and \rxte PCA lightcurves binned  
at 128~s. The \hetg lightcurve was made from combined
0.8--7 keV HEG and MEG data
(first order only).
The PCA lightcurve was made from PCU2 only,
in the 3--13 keV band (the upper energy was chosen
to be less than 15 keV as an extra precaution
against background contamination). 
It can be seen that the \chandra lightcurve
covers $\sim 2$ days, with a large gap of $\sim 1/3$ day
starting $\sim 30$~ks into the observation. The PCA lightcurve 
intermittently covers all of the period of the \chandra
observation.
The large amplitude, rapid variability which is
characteristic of \src is again evident in the
lightcurves. Variations by a factor $\sim 2$ on timescales
of hundreds of seconds occur frequently, and a very large amplitude
drop in flux is seen toward the end of the 
observations, when the flux dropped by a factor $\sim 4$ in
$\sim 3$~ks.

We extracted intensity-selected HEG, MEG, and PCA
spectra such that `low-state' spectra were made from
time intervals during which the 0.8--7~keV 
first order \hetg (i.e. HEG plus MEG) count rate
was less than 0.60 ct/s (dotted line in \figlcurvesp),
and `high-state' spectra were made from intervals
with $>0.61$ \hetg first-order ct/s.
This resulted in a low-state spectrum which
had $0.1580 \pm 0.0016$ HEG ct/s, $0.2811 \pm 0.0022$ MEG ct/s, 
and a net exposure
time of 59.440~ks. The corresponding values for the
high state were 
$0.2610 \pm 0.0028$ HEG ct/s, $0.4796 \pm 0.0038$ MEG ct/s, and 
32.869~ks net exposure time. 
Since the overlap of \chandra and \rxte is not perfect,
this resulted in low-state and high-state PCA
spectra which had smaller exposure times than the
corresponding \chandra spectra. We obtained
exposure times of 33.712~ks and 15.888~ks for the low-state
and high-state PCA spectra respectively. Strictly
speaking this does not sample the same time intervals
as \chandra so we shall bear this in mind when interpreting
the \rxte data. The implicit assumption is that those
parts of the low and high states `missed' by \rxte are,
on average,
spectrally similar to the parts that are not missed. 
The mean 3--15 keV background-subtracted 
count rates for the combined PCU0 and PCU2 spectra
were $2.709 \pm 0.011$ ct/s and $3.971 \pm 0.017$ ct/s
for the low-state and high-state respectively.

The time intervals derived from the \chandra intensity selection
consist of alternating low and high states.
The longest interval between those alternate
states was 12.544~ks (not including the large $\sim 1/3$ day gap in the
data -- see \figlcurvesp). Thus, if we find  variability
in spectral features observed by {\it Chandra}, that variability would be occurring
on a timescale of $<12.544$~ks, since the 
mean count rate, lightcurves, and spectra 
are similar either side of the gap.
If the mass of the central black hole in NGC~7314 is
$5 \alpha \times 10^{6} \ M_{\odot}$ (where $\alpha \sim 1$ -- see \S\ref{intro}) 
then 12.544~ks corresponds to
a light-crossing distance of $\sim 502 r_{g}/\alpha$ and
a Keplerian orbital radius of $\sim 18.5 \alpha^{-\frac{2}{3}} \ r_{g}$ (where $r_{g}
\equiv GM/c^{2}$).

\section{HIGH-RESOLUTION, TIME-RESOLVED SPECTROSCOPY}
\label{hegspec}

Below we describe spectral fitting of the mean, low-state
and high-state spectra, using XSPEC v11.2 (Arnaud 1996).
The $C$-statistic was used for minimization and,
unless otherwise stated, all statistical errors 
correspond to 90\% confidence for one interesting parameter
($\Delta C = 2.706$). 
All model parameters will be
referred to the source frame using $z=0.004760$,
unless otherwise stated. As mentioned in \S\ref{intro}, this 
value of redshift was obtained
by Mathewson \& Ford (1996) using H~{\sc i}  measurements
so is probably the most accurate available. The range of
measurements of $z$ in NED\footnote{NASA Extragalactic
Database} correspond to an uncertainty of
less than 10~$\rm km \ s^{-1}$, which is much less
than the systematic uncertainty in the \hetg 
wavelength scale 
($\sim 433 \ \rm km \ s^{-1}$ ($\sim 11$~eV) at 6.4 keV for the 
HEG\footnote{http://space.mit.edu/CXC/calib/hetgcal.html}). 

The MEG 0.5--5~keV mean spectrum can be described
by an absorbed power law with photon index $1.99 \pm 0.06$
and column density $10.2^{+0.6}_{-0.5} \times \rm \ 10^{21} \ cm^{-2}$,
plus $(4.3 \pm 0.5)\%$ of the direct, unabsorbed component of the power-law 
continuum. 
Similar spectral fitting of the low-state and high-state spectra
reveal no evidence of variability in the column density, power-law slope,
or scattering fraction. However, the power-law slope is not well-constrained
and we will show below (\S\ref{pcafits}) that the \rxte
data do show variability in the slope.
Hereafter, for {\it Chandra},
we will discuss only the HEG spectra since 
we are primarily interested in the Fe-K region, and the
energy resolution and effective area are superior to the
MEG at high energies.

\subsection{Rapid Variability of the Fe-K Complex}
\label{fekvariability}

\fighegfek (a) shows the low-state HEG spectrum in the 5.5--8.5 keV
observed energy range. \fighegfek (b) and \fighegfek (c) show
the HEG high-state and mean spectra, respectively.
Dotted lines show the expected positions of
the Fe~{\sc i}~$K\alpha$ line, the Fe~{\sc xxv}
He-like triplet, and the Fe~{\sc xxvi}~Ly$\alpha$ and
Ly$\beta$ lines.
In the low-state spectrum shown in \fighegfek (a) there are actually
five discrete emission-line features
immediately apparent in the 5.5--8.5~keV band.
One is at an energy consistent
with Fe~{\sc i}~$K\alpha$ at 6.400~keV in the source frame, although 
the energies of $K\alpha$ lines from all ions up
to Fe~{\sc xvii} differ from 6.400~keV by less
than the \chandra HEG resolution (FWHM~$\sim 39$~eV).
Hereafter, we shall refer to this line as \fekalfa (I).
Another emission line in \fighegfek (a) is 
near the expected energy for the He-like forbidden
line but redshifted, and a third is near that expected for \feklyap, 
again redshifted. Although
this redshift is small ($\sim 35$~eV), it is
significant. 
The systematic uncertainty in the
energy scale is a factor of $\sim 3$ less than
the redshift of the \fexxv and \fexxvi $K\alpha$ lines. 
The \fekalfa (I) line is not redshifted relative to 6.400~keV.
However we do not know the dominant ionization stage of Fe
producing the \fekalfa (I) line, so a redshift consistent with
the \fexxv and \fexxvi $K\alpha$ lines is not ruled out.

\figcontours shows two-parameter joint
confidence contours for intensity of line emission versus
energy for the low-state, mean, and high-state spectra
in the 6.1--7.1~keV band. The confidence levels correspond
to 68\% (blue), 90\% (red), and 99\% (black).
The contours were generated by fitting a simple power-law
continuum in the 5--9 keV band (see below for details)
and constructing a grid of $\Delta C$ values when a
narrow ($\sigma = 1$~eV) Gaussian component is added to the model,
whose energy and intensity were traced over the range
6.1--7.1~keV and 0--$6 \times 10^{-5} \rm \ photons \ cm^{-2}
\ s^{-1}$ respectively. Here, and in subsequent spectral
fitting, the low-state and high-state spectra were 
binned at $0.01\AA$ and the mean spectrum was binned
at $0.005\AA$ (the HEG FWHM resolution is $0.012\AA$).

In the low state there are
three clear and isolated, discrete emission features
in the 6.1--7.1~keV band, corresponding to $K\alpha$
emission from Fe~{\sc i}--Fe~{\sc xvii}, \fexxvp, and \fexxvip.
\fighegfek and \figcontours clearly show that the
complex of Fe-K emission, and therefore the ionization balance
of Fe, is rapidly variable on timescales $<12.5$~ks.
It appears that as the continuum level increases
(by a factor of $\sim 1.7$ from low-state to high-state), the
ionization structure of the emitting region changes to
produce a complex and variable blend of emission,
instead of three, simple, discrete emission lines.
Note that the signal-to-noise of the low-state and high-state
spectra is approximately the same, and the signal-to-noise of the
mean spectrum is approximately twice that of either low-state or high-state
spectra alone. Therefore the differences between the
contours 
in different states (\figcontoursp) are {\it not} 
due to signal-to-noise effects. Nor are the differences simply
the result of line emission becoming less significant against
a rising continuum. \figcontours shows that the {\it centroids and
widths} of the line emission near 6.4~keV and 6.6~keV appears
to vary. It also appears that the
\feklya emission becomes broader.
Detailed spectral fitting (described below) shows
marginal evidence that the \feklya centroid
energy is higher by $\sim 10$~eV
in the mean spectrum relative
to the low-state.
The intensity of \feklya appears to be consistent
with a constant in all three spectra (see \fighegfekp, \figcontoursp, and
\tablehegfitsp). However, the behavior
of the intensity of this line in response to
continuum variations is not expected to be simple because as the
continuum level increases, more Fe~{\sc xxv} may get ionized
to Fe~{\sc xxvi}, but at the same time more Fe would become
completely stripped.
Below we examine the low-state, mean, and high-state spectra
in more detail. 

\subsection{Detailed HEG Spectra Around the Fe-K Complex}
\label{lowstate}

\fighegfek (a) shows that there is
a tentative detection of \feklyb which
has a redshift that is consistent with \feklya (details in \S\ref{felybeta}). 
There is also a strong emission line which is
unidentified (it is at $\sim 5.84$~keV in the
source frame). There are no candidate atomic features
near 5.84~keV which have high enough oscillator strength
and elemental abundance to produce the line, and yet  
do not also predict other features which should be 
stronger but are not observed.
We can certainly eliminate any instrumental
artifact being responsible because the line is
almost undetected in the
high-state spectrum (\fighegfek [b]) and
weak in the mean spectrum (\fighegfek [c]). 
It is also not a background feature: the background
spectrum, extracted from 3.6--10 arcsec either side
of the dispersion axis is shown plotted in green in \fighegfek (a).
The most likely origin of the $\sim 5.84$~keV feature is
\fekalfa emission which has
suffered gravitational and Doppler
energy shifts in a spatially localized region
near the central black hole, and this is discussed in detail
in \S\ref{narroworigin}.

Intially, we fitted the four strongest emission lines
(i.e. all except the \feklyb line)
in the low-state spectrum,
with Gaussians. By fitting the continuum 
in two different ways, we demonstrate,
for the low-state spectrum, that the derived
line parameters are robust to different
assumptions about how the continuum is modeled
since the lines are clearly narrow compared to the
spectral resolution. In the first method we simply fitted
a power law in the 5--9 keV range
(as well as the four Gaussians). 
We found that
extrapolation of this power law back down to 3 keV
still provided an excellent description of the data.
In the second method, we fitted the 0.8--9 keV data
with an absorbed power law (omitting the 2--2.5~keV
band due to systematic residuals around sharp changes 
in the effective area of the telescope --
see Yaqoob \etal 2003), plus the four Gaussians.
We then used the best-fitting
photon index and ignored the data below
5 keV. 
In both cases the photon index was frozen
to obtain the statistical errors on the Gaussian
parameters. The intrinsic widths of the Gaussians
were fixed at 1~eV (i.e. much less
than the spectral resolution), but we allowed them to float
one at a time in order to obtain upper limits.
All four emission lines are unresolved by the HEG.
Full results are shown in \tablehegfitsp.
It can be seen from \tablehegfits
that the line parameters for the two
continuum forms used are either identical or indistinguishable.
Therefore we simply used the first method
(simple power law fitted over 5--9~keV)
for modeling the continuum for the mean and high-state
spectra. Full results 
of the continuum plus line fits to the
mean and high-state spectra are shown in \tablehegfitsp.
Note that the best-fitting model
to the low-state spectrum (first method) is shown
as a red solid line in {\it all} the panels of \fighegfekp.

We estimated the significance of each of the four detected
emission lines in the low state by removing each emission line from the best-fitting
four-Gaussian
model and then re-fitting the model, noting the increase
in the $C$-statistic. The best-fitting four-Gaussian
model was then replaced before investigating another line
in the same manner.
The 5.842~keV line was detected only with
$\sim 99\%$ confidence, and its equivalent width (EW)
is $32^{+64}_{-16}$~eV. The \fekalfa (I) line was
detected with a confidence level $> 4\sigma$, and the
$K\alpha$ lines from \fexxv, and \fexxvi were detected
at $> 3\sigma$ confidence.
The EWs of \fekalfa (I), \fexxv (f), and \feklya in
the low-state spectra are
$81^{+34}_{-34}$, $59^{+25}_{-34}$, and $68^{+53}_{-30}$~eV 
respectively (see \tablehegfitsp).
The $K\beta$ line corresponding to the \fekalfa (I) line is not
detected, but by adding another narrow Gaussian at 7.058~keV
(source frame) to the best-fitting, four-Gaussian,
low-state model, we obtained upper limits on the intensity and
EW of the Fe~$K\beta$ line of $1.6 \times 10^{-5}
\ \rm photons \ cm^{-2} \ s^{-1}$, and 62~eV respectively,
using the low-state spectrum.
This is consistent with the Fe~{\sc i}~$K\beta$:$K\alpha$
branching ratio of 17:150 (e.g. Bambynek 1972).

\begin{table*}[tbh]
\caption{\chandra HEG Spectral Fitting Results}
\begin{tabular*}{\textwidth}{@{}l@{\extracolsep{\fill}}rrrr}
\hline
& & & & \\
Parameter & Low State (I) $^{a}$ & Low State (II) $^{a}$ &
Mean & High State \\
& & & & \\
\hline
& & & & \\

$C$ (degrees of freedom) &
117.7 (99) &   118.7(100) & 236.9/(211) & 116.6/(103) \\

{\bf Fe~K$\alpha$ (I) $^{b}$} & & & & \\

$E_{0}$ (keV) & $6.405^{+0.016}_{-0.017}$ & $6.405^{+0.016}_{-0.017}$
        & $6.412^{+0.010}_{-0.016}$ & fixed (6.405) \\

$1-(E_{0}/6.400)$ & $-0.8^{+2.7}_{-2.5} \times 10^{-3}$
        & $-0.8^{+2.7}_{-2.5} \times 10^{-3}$ & $-1.9^{+2.5}_{-1.6} \times 10^{-3}$
        & \ldots  \\

$c \Delta E/6.400$ ($\rm km \ s^{-1}$) & $-240^{+810}_{-750}$
        & $-240^{+810}_{-750}$ & $-570^{+750}_{-480}$ & \ldots \\

$\sigma$ (keV) & $<0.032$ & $<0.032$ & \ldots  & \ldots \\

FWHM ($\rm km \ s^{-1}$) & $<3520$ & $<3520$ & \ldots & \ldots \\

$I \ \rm (10^{-5} \ photons \ cm^{-2} \ s^{-1})$
        & $2.4^{+1.0}_{-1.0}$ & $2.5^{+0.9}_{-1.1}$
        & $1.7^{+0.7}_{-0.8}$ & $0.8^{+1.3}_{-0.8}$ \\

EW (eV) & $81^{+34}_{-34}$ & $83^{+30}_{-37}$ & $48^{+20}_{-23}$
        & $18^{+29}_{-18}$ \\

{\bf Fe~K$\alpha$ (II)} $^{b}$ & & & & \\

$\Delta C$ $^{c}$ & 0.0 & $-0.3$ & $-0.2$ & $-6.5$ \\

$E_{0}$ (keV) & fixed (6.325) & fixed (6.325) & fixed (6.325)
                & $6.325^{+0.026}_{-0.014}$ \\

$1-(E_{0}/6.400)$ & \ldots & \ldots & \ldots  & $1.17^{+0.22}_{-0.41} \times 10^{-2}$ \\

$c \Delta E/6.400$ ($\rm km \ s^{-1}$) & \ldots & \ldots & \ldots  &
        $3510^{+660}_{-1230}$ \\

$\sigma$ (keV) & \ldots & \ldots & \ldots  & $<0.040$ \\

FWHM ($\rm km \ s^{-1}$) & \ldots & \ldots & \ldots & $<4460$ \\

$I \ \rm (10^{-5} \ photons \ cm^{-2} \ s^{-1})$
        & $<0.3$ & $<0.3$ & $<0.7$ & $1.8^{+1.5}_{-1.3}$ \\

EW (eV) & $<7$ & $<7$ & $<15$ & $39^{+33}_{-28}$ \\

{\bf Fe~{\sc xxv} (f)} & & & & \\

$E_{0}$ (keV) & $6.607^{+0.011}_{-0.017}$ & $6.607^{+0.011}_{-0.017}$
        & fixed (6.607) & fixed (6.607) \\

$1-(E_{0}/6.637)$ & $4.5^{+2.6}_{-1.7} \times 10^{-3}$
        & $4.5^{+2.6}_{-1.7} \times 10^{-3}$  & \ldots & \ldots \\

$c \Delta E/6.637$ ($\rm km \ s^{-1}$) & $1350^{+780}_{-510}$
        & $1350^{+780}_{-510}$ &  \ldots & \ldots \\

$\sigma$ (keV) & $<0.061$ & $<0.061$  &  \ldots & \ldots \\

FWHM ($\rm km \ s^{-1}$) & $<6510$ & $<6510$  &  \ldots & \ldots \\

$I \ \rm (10^{-5} \ photons \ cm^{-2} \ s^{-1})$
        & $1.9^{+0.8}_{-1.1}$ & $1.9^{+0.8}_{-1.1}$ &
        $1.1^{+0.7}_{-0.6}$ & $0.2^{+1.3}_{-0.2}$ \\

EW (eV) & $59^{+25}_{-34}$ & $60^{+25}_{-35}$  &
        $29^{+20}_{-17}$ & $3^{+28}_{-3}$ \\


{\bf Fe~{\sc xxvi}~Ly$\alpha$ } & & & &  \\

$E_{0}$ (keV) & $6.931^{+0.018}_{-0.011}$ & $6.931^{+0.018}_{-0.011}$
        & $6.940^{+0.010}_{-0.010}$ &    fixed (6.931) \\

$1-(E_{0}/6.966)$ & $5.0^{+1.6}_{-2.6} \times 10^{-3}$ &
        $5.0^{+1.6}_{-2.6} \times 10^{-3}$ &
        $3.7^{+1.5}_{-1.4} \times 10^{-3}$ & \ldots \\

$c \Delta E/6.966$ ($\rm km \ s^{-1}$) & $1500^{+480}_{-780}$ & $1500^{+480}_{-780}$
        & $1110^{+450}_{-420}$ & \ldots \\

$\sigma$ (keV) & $<0.030$ &  $<0.030$ & $<0.026$ & $<0.101$ \\

FWHM ($\rm km \ s^{-1}$) & $<3050$ & $<3050$  & $<2640$ & $<10270$  \\

$I \ \rm (10^{-5} \ photons \ cm^{-2} \ s^{-1})$
        & $1.8^{+1.4}_{-0.8}$ & $1.9^{+1.4}_{-0.8}$ & $1.9^{+1.0}_{-0.7}$
        & $1.7^{+1.8}_{-1.2}$ \\
EW (eV) & $68^{+53}_{-30}$ & $71^{+53}_{-30}$ & $63^{+33}_{-23}$
        & $44^{+47}_{-31}$ \\
\end{tabular*}
\end{table*}
\setcounter{table}{0}
\begin{table*}[tbh]
\caption{--{\it Continued}}
\begin{tabular*}{\textwidth}{@{}l@{\extracolsep{\fill}}rrrr}
\hline
& & & & \\
Parameter & Low State (I) $^{a}$ & Low State (II) $^{a}$ &
Mean & High State \\
& & & & \\
\hline
& & & & \\

{\bf Unidentified}  & & & & \\

$E_{0}$ (keV) & $5.842^{+0.12}_{-0.14}$ & $5.842^{+0.12}_{-0.14}$
        & fixed (5.842) & fixed (5.842) \\

$\sigma$ (keV) & $<0.030$ &  $<0.030$ & \ldots & \ldots \\

FWHM ($\rm km \ s^{-1}$) & $<3620$ & $<3620$ & \ldots & \ldots \\

$I \ \rm (10^{-5} \ photons \ cm^{-2} \ s^{-1})$
        & $1.0^{+2.0}_{-0.5}$ & $1.0^{+2.0}_{-0.5}$
        & $0.7^{+0.6}_{-0.5}$ & $0.4^{+1.0}_{-0.4}$ \\

EW (eV) & $32^{+64}_{-16}$ & $32^{+64}_{-16}$
        & $19^{+16}_{-14}$ & $8^{+20}_{-8}$ \\

& & & & \\

F(2--10 keV) $^{d}$ & 2.5 & 2.5 & 3.0 & 4.0 \\

L(2--10 keV) $^{e}$ &  1.2 & 1.2 & 1.5 & 2.0 \\

& & & & \\

\hline
\end{tabular*}
{\small Gaussian modeling of the emission lines in
the NGC~7314 HEG data (see \S\ref{hegspec}).
The baseline model has four Gaussians (see \fighegfek and \S\ref{hegspec}).
The intrinsic width of each Gaussian is fixed
at 1~eV, but allowed to float in turn to determine the upper limit
(temporarily freezing the line energy in the process).
Statistical errors and upper limits are for 90\% confidence
($\Delta C = 2.706$). All emission-line parameters refer to
the rest-frame of NGC~7314 ($z=0.004760$). Velocities have been
rounded to the nearest 10~$\rm km \ s^{-1}$. \\
$^{a}$ All except the low-state (II) results refer
to fits with a simple power law for the continuum fitted in the
5--9 keV range. To demonstrate the robustness to details of how
the continuum is modeled, the low-state (II) results are
obtained by fitting the continuum with an
absorbed power law in the 0.8--9 keV band. See \S\ref{hegspec} for details. \\
$^{b}$ Two distinct \fekalfa components are detected. \fekalfa (I) is
not variable and is
not redshifted relative to 6.400 keV and
is therefore a `distant-matter' component. \fekalfa (II) {\it is}
variable and is redshifted relative to 6.400 keV. \\
$^{c}$ Change in $C$-statistic when the
\fekalfa (II) component is added to the four-Gaussian model. \\
$^{d}$ {\it Observed} flux in units of $\rm 10^{-11} \ ergs \ cm^{-2} \ s^{-1}$. \\
$^{e}$ {\it Intrinsic} luminosity in
the source frame, in units of $10^{42} \rm \ ergs \ s^{-1}$,
assuming $H_{0} = 70 \rm \ km \ s^{-1} \ Mpc^{-1}$, and $q_{0}=0$.}
\end{table*}

\fighegfek~(b) and \figcontours show that in the high state
the line at 6.4~keV is hardly detected, but
a new emission feature appears at $\sim 6.3$~keV.
However, it is detected only at $\sim 96\%$ confidence. Nevertheless, for
completeness, we model it with an additional Gaussian component and
include the results in \tablehegfits and refer to the
feature as \fekalfa (II). 

\subsubsection{The \fexxv He-like Triplet}
\label{hetriplet}

The Fe~{\sc xxv} He-like triplet is potentially
a powerful diagnostic of density and temperature. However,
we note that the separation of the forbidden and intercombination
lines and the intercombination and resonance lines is only
31.2 and 14.5~eV respectively \footnote{We take the energies
of the forbidden (f), intercombination (i), and resonance lines (r) from
NIST ({\tt http://physics.nist.gov/PhysRefData/}):
6.6365, 6.6677, and 6.6822 keV respectively.}, less than the HEG FWHM (43~eV) at
the observed energy of the intercombination line. 
We {\it assume} that the observed \fexxv line is principally the forbidden line,
since the measured
redshift of $1350^{+780}_{-510} \ \rm km \ s^{-1}$ (relative to the systemic velocity of
NGC~7314) is then consistent with the
redshift of $1500^{+480}_{-780} \ \rm km \ s^{-1}$
for \feklya (see \tablehegfitsp).
For \feklya we used a rest energy of 6.966~keV: a branching ratio of 2:1 for the
Ly$\alpha_{1}$ ($\lambda 1.778$) and Ly$\alpha_{2}$ ($\lambda 1.783$)
components gives a centroid energy of 6.966 keV (Pike \etal 1996).

If the \fexxv line were the intercombination or resonance line,
the redshifts would be 
$2730 \ \rm km \ s^{-1}$ and $3380 \ \rm km \ s^{-1}$
respectively.
However, we note that the upper limit on the width
of the \fexxv line (low state) is about twice that of \feklyap, although
they have similar EW (see \tablehegfitsp).
Until {\it Astro-E2} resolves the He-like Fe triplet, we will
assume it is dominated by the forbidden line.
For the forbidden line to dominate,
the electron density should be less than $\sim 10^{16} \ \rm cm^{-3}$
(e.g. Bautista \& Kallman 2000). 

We obtained upper limits on the
intensities of the intercombination (i) and resonance (r) lines of 
0.6 and 0.7~$\times 10^{-5} \ \rm photons \ cm^{-2} \ s^{-1}$
respectively, by adding narrow Gaussians, one
at a time, at energies redshifted from systemic by the
same redshift as \feklyap.
Then, taking the lowest intensity in the measured
range (low state) of what we assume is the forbidden line, 
we obtained lower limits on $R \equiv \rm f/i$
and $G \equiv \rm (f+i)/r$ of 1.3 and 1.1 respectively.
As Bautista \& Kallman (2000) point out, 
caution should be exercised with the use of $R$ and $G$ 
as density and temperature diagnostics respectively, since
they are 
sensitive to whether the plasma is collisionally ionized or photoionized.
Also, there are many dielectronic satellite features from He-like
Fe in the range 6.61--6.71~keV (Oelgoetz \& Pradhan 2001).
Nevertheless, the density curves in Bautista \& Kallman (2000) are
consistent with $n_{e} < 10^{16} \ \rm cm^{-3}$ and
the temperature curves indicate that $T<10^{6.6}$~K or
$T<10^{7.6}$~K for collisional and photoionized gas respectively.
Note that even the higher temperature corresponds to 
a thermal Doppler width of only $\sim 100 \ \rm km \ s^{-1}$,
much less than the HEG resolution, but is $\sim 37\%$ of the
6~eV FWHM resolution of the \astroe calorimeter at 6.7~keV.

\subsubsection{\fekalfa Line Emission from Distant Matter?}
\label{energyshifts}

For the \fekalfa (I) line,
the velocities shown in \tablehegfits assume an origin in
Fe~{\sc i}, for which the
branching ratio of 2:1 for the 
$K\alpha_{1}$ ($\lambda 1.936$) and $K\alpha_{2}$ ($\lambda 1.940$)
components gives a centroid energy of 6.400 keV (Bambynek \etal 1972).
Ionization states up to Fe~{\sc xvii} will give
a $K\alpha$ line centered at an energy which 
differs from 
6.400~keV by less than the HEG resolution (FWHM~$\sim 39$~eV at 6.4~keV).
It can be seen (\tablehegfitsp) that the \fekalfa (I) line 
energy shift 
is consistent with zero, but obviously there
is some uncertainty as to the dominant ions
responsible for the emission. 

Given the tentative rapid variability of the line emission near 6.4~keV,
it is not clear whether there is a contribution to the
line emission at 6.4~keV from distant matter. 
If the low-state \fekalfa~(I) line were predominantly from 
distant matter, one could estimate a lower limit to the
distance of the emitter from an upper limit on the
width of the line emission.
Assuming a virial relation and an r.m.s.
velocity of
$\sqrt{3}V_{\rm FWHM}/2$ (e.g. Netzer 1990), and using
our measured upper limit of $\rm FWHM< 3520 \ \rm
km \ s^{-1}$ (\tablehegfitsp), 
gives a location
in the outer BLR or beyond, at $>2.8\alpha$ light days from the central mass
($5\alpha \times 10^{6} \ M_{\odot}$).

The amount of Fe in the line-of-sight is not sufficient
to produce the observed EW of the
\fekalfa (I) line ($81 \pm 34$ eV) and this
further argues for the  \fekalfa~(I) line originating
close to the X-ray source, possibly in the accretion disk. The Fe-K edge
optical depth ($\tau_{\rm Fe} = 0.018$) inferred from the soft X-ray cut-off
(neutral Hydrogen column density of $10^{22} \ \rm cm^{-2}$ -- see \S\ref{hegspec})
can at most produce an EW of 8.5 eV, 
even if it fully covered the sky (e.g. see Krolik \& Kallman 1987; Yaqoob \etal 2001).
On the other hand, the upper limit on the 
threshold optical depth of an Fe~{\sc i} K-edge, 
obtained from directly from fitting the HEG data with
an edge model at 7.11~keV (rest-frame), is $\tau_{\rm Fe} =
0.12^{+0.14}_{-0.12}$. 

\subsubsection{Diagnostics from the Fe Ly$\beta$:Ly$\alpha$ Ratio.}
\label{felybeta}

If \feklya is observed then \feklyb emission is
{\it expected}, with a predictable relative
intensity (which may be zero), from basic physics.
The theoretical
value of the Fe Ly$\beta$:Ly$\alpha$
ratio depends on the excitation mechanism
for the \feklya line (e.g. see Bautista \etal 1998; Bautista \& Titarchuk 1999).
\feklya emission directly following
recombination will be accompanied by an edge at 9.19~keV
and no Ly$\beta$ emission. The Ly$\beta$:Ly$\alpha$ ratio
for collisional excitation depends on temperature and
is potentially a powerful temperature diagnostic.
At low temperatures, the Ly$\beta$:Ly$\alpha$ ratio
can be negligible, but it can be as high as 0.44
for $kT>1$~keV. In the case of resonant fluorescent excitation
the Ly$\beta$:Ly$\alpha$ ratio depends on the intensity
and shape of the radiation field and may be as high as 0.26.
It should therefore be clear that measurement of,
or placing constraints on, the Fe Ly$\beta$:Ly$\alpha$ ratio 
is critically important even if \feklyb is not detected.
This is especially true since some parts of parameter
space predict Ly$\beta$:Ly$\alpha \sim 0$.

Although the detection of \feklyb in our data is only
marginal (at best, for the low-state spectrum,
$\Delta C =4.5$ for the addition of
a Gaussian with two free parameters), we can 
measure the center energy and intensity. We obtain 
$E = 8.18^{+0.03}_{-0.04}$~keV, an intensity of
$1.0^{+2.7}_{-0.9} \times 10^{-5} \ \rm photons \ cm^{-2} \ s^{-1}$,
and an EW of $59^{+159}_{-53}$~eV
(all are source-frame quantities, measured from the low-state
spectrum). The redshift relative
to the expected energy of 8.250~keV \footnote{We take the wavelengths
of $1.5024\AA$ and $1.5035\AA$ for Fe~{\sc xxvi}~Ly$\beta_{1}$
and Ly$\beta_{2}$ respectively, from Verner, Verner, \& Ferland (1996).
The branching ratio of 2:1 for Ly$\beta_{1}$:Ly$\beta_{2}$
gives a centroid energy of 8.250~keV} 
is $8.5^{+4.6}_{-3.7} \times 10^{-2}$, consistent with the redshift we
measure for \feklya of $5.2^{+1.4}_{-1.5} \times 10^{-2}$. 
Our absolute lower limit on the ratio of \fexxvi Ly$\beta$ to
Ly$\alpha$ intensities is 0.05. 
Our measurement errors on the Ly$\beta$:Ly$\alpha$ ratio
are obviously too large for this to be a useful diagnostic yet, but it is
something future missions will very soon begin to
address, yielding critical information from a simple measurement.

\subsubsection{Absorption Lines in the High-State Spectrum}
\label{highmeanstate}

\fighegfek~(b) shows that the high-state spectrum has
two narrow absorption features that are not present
in the low-state and mean spectra.
Using an inverted Gaussian added to the best-fitting
high-state model (\tablehegfitsp) we measure
rest-frame energies of
$6.000^{+0.012}_{-0.034}$~keV and
$6.760^{+0.018}_{-0.011}$~keV, and corresponding equivalent
widths of $16.5^{+3.3}_{-7.7}$~eV and $16.6^{+4.4}_{-8.9}$~eV
respectively. The origin of 
neither is clear, since the energies do not correspond to any known
atomic transitions which predict the appropriate absorption,
without predicting other features which are not observed.
If the  features are due to 
\fexxv resonance absorption, the lower energy line
implies an inflow of $\sim 0.10c$, whilst the higher energy
feature implies an outflow of $\sim 0.012c$. Although
the statistical significance of either of the features is not
high (99\% and 97\% confidence for the lower and higher energy
line respectively), the physical implications are
important enough to warrant attempting confirmation with future
missions (especially for the lower energy line since
evidence for inflow is so rare). Moreover, unambiguous 
narrow Fe-K absorption
features indicating velocities as high as $\sim 0.1c$
have been found by \xmm in the NLS1 PG~$1211+143$
(Pounds \etal 2003).

\section{PHYSICAL CONSTRAINTS FROM THE CHANDRA DATA}
\label{releffects}

\subsection{Geometry}
\label{geometry}

While the redshifts of the \fexxv~(f),
and \feklyb lines are tentative, the redshift of
\feklya is firm, and is a robust measurement.
Since \feklya is not resolved by the HEG, the 
upper limit on its width (\tablehegfitsp) is also
a robust, model-independent measurement.
The redshift and width constraints on the \feklya 
line cannot both be simultaneously satisfied if the
line were produced in a fully covering spherical geometry.
For spherical inflow or outflow
one would expect a P Cygni line profile which is not observed.
We quantified the possibility of a P Cygni line profile
in the low-state data (where the lines are the clearest),
by adding an inverted Gaussian to the continuum plus
four-Gaussian line model (see \S\ref{lowstate}). The
inverted Gaussian (absorption line) was blueshifted from
6.966~keV by the same amount as the observed \feklya line
is redshifted from 6.966~keV. We obtained a best-fitting
equivalent width of zero and a one-parameter, 90\% upper limit
of 14~eV. This is to be compared with the measured
EW of $68_{-30}^{+53}$~eV for the \feklya
emission line (\tablehegfitsp). If the He-like line is
the forbidden or intercombination component, it could not of course have
a P Cygni line profile. If the He-like line is
the red part of a P Cygni line profile
from the resonance transition, the emission should 
extend all the way to the rest energy but it
clearly does not (\fighegfek and \figcontoursp). 
We conclude that the data do not support
a P Cygni line profile for the He-like and H-like lines.
 
In the case of Keplerian orbits in a spherical distribution, 
if the line-emitting matter were close enough to produce the
required redshift and satisfy the time-variability constraints, 
the line profile would become broader than observed.  
A partially covering (or partially obscured) spherical geometry 
or cone model (e.g. see Zheng, Sulentic, \& Binette 1990;
Sulentic \etal 1998), or anisotropic illumination model
(Yaqoob \etal 1993) would involve
very restricted solid angles (to satisfy
the constraint that the lines are not resolved
by \chandra). Moreover, the structure or illuminating cone must be one-sided
to give only a redshift and not a blueshift as well.
To be sure, a bi-polar inflow or outflow, or wind, in which the
approaching side is obscured, cannot be ruled out.
For example, if the flow were aligned with the rotation
axis of an accretion disk, and the linear extent of
the flow were smaller than the disk radius, and if
the system were viewed close to face-on with the flow
facing away, then the approaching side of the flow
would be hidden from view.
However, the solid angle subtended 
at the continuum X-ray source by such a flow would be too small to
produce the fairly large EW of \feklya ($\sim 70$~eV in
the low-state spectrum and $\sim 60$~eV in the mean spectrum).
It is also difficult to explain the line redshift and 
width constraints in terms of the wind outflow model of Elvis (2000)
without further fine-tuning of the model.
To overcome the blueshift from the outflow
(in order to give an observed redshift), the wind 
must be placed closer to the black hole in order for
gravitational redshifts to compensate, but then the lines would be
broader than observed.

It is conceivable that the the 5.842~keV and 6.931~keV lines could be 
from a highly collimated jet (bulk velocity, $> 0.055c$),
seen nearly end-on, since these
line are approximately equidistant from 6.4~keV? The small
difference of 27~eV would then be due to a slight
difference in the angles of the receding and approaching sides
of the jet relative to the line of sight. This scenario
is unlikely because the narrow line widths require a small
dispersion in the bulk velocity of $\sim 10\%$ and the scenario does not
account for the 6.607~keV line. 

All of the above models must also
account for the rapid response ($<12.5$~ks) of the ionization balance of Fe 
to the X-ray continuum variability. Given the amount of fine-tuning and/or
incompatibility with the data, we consider the most viable and least-contrived
origin of the line emission from highly ionized Fe to be the accretion disk
itself.

\subsection{The Origin of Narrow Lines from Disks}
\label{narroworigin}

If the unresolved narrow emission lines
observed by \hetg are from a disk
there are three possible reasons for them to be narrow,
which we summarize below.

(1) The lines could be from  individual `hot spots' on the
disk, whose orbital period is less than the length
of the observation. The line width is then determined
by the arc length traveled by the rotating hot spot 
during the observed time, and by the spread in 
radial coordinate.

(2) A given observed line could be the red or blue Doppler
horn of a much wider line profile, originating in 
a ring of emission from either a single hot spot observed
for more than one orbit, or a collection of hot spots
observed for less than one orbit that are
distributed in the form of a ring. The ring cannot
have too large of a spread in radius because emission
from different radii will effectively broaden the
red and blue horns and therefore the observed line.
MHD simulations by  Armitage and Reynolds (2003)
show that hot spots can survive for several
orbits and are capable of producing sharp `spikes' such as
those observed here.

(3) The {\it entire} line profile may be 
from a large area on the disk but it may still
so narrow
that it is unresolved even by Chandra.
This could happen for disks which have a flat radial
line emissivity and are near face-on.
In this case, we are no longer restricted to hot spots
or annuli of emission since the red and blue
Doppler horns are not resolved, so the emission could
even extend down to small radii, if the radial emissivity
law is flatter than $r^{-2}$. 

In the case of (1) 
and (2), for a given disk inclination angle,
there is only a very restricted choice in the radial coordinate
($r/r_{g}$)
in order to match the observed line energies
with the predicted energies. Therefore, it would have to be
a remarkable coincidence for the observed energies of
6.607~keV and 6.931~keV (\tablehegfitsp) to lie so close
(within only
$\sim 1500 \rm \ km \ s^{-1}$) to the rest energies
of \fexxv (f) and \feklya respectively. The extent of the
fine tuning required by scenarios (1) and (2) is illustrated
in \fighornsp, which shows the relation between the
disk inclination angle, $\theta$ (for the extreme envelopes
$\theta=0^{\circ}$ and $\theta=90^{\circ}$), and the radial coordinate,
for a given rest energy of the line. Blue and red curves
indicate emission from approaching and receding sides of the
disk. Horizontal lines in the panels on the left are
shown for line rest energies of $E_{0} = 6.400$, 6.636, and 6.966~keV.
Each pair of panels in \fighorns corresponds to the 
energy of an observed line (6.931, 6.607, and 5.842~keV, as indicated).
A solution exists where lines of a given $E_{0}$ would intersect 
blue and red curves of a given $\theta$. This gives $r/r_{g}$ and
also indicates that the energy range of the whole line profile
covers the vertical interval in $E_{\rm obs}/E_{0}$ at that
$r/r_{g}$, which is contained inside blue and red curves
corresponding to the correct disk inclination angle.
This inclination angle cannot be determined 
uniquely without measuring both red and blue Doppler horns,
as the right-hand panels of \fighorns show
(although the plots also show that there are some restrictions on $\theta$).
Note that we would not necessarily be able to observe both
red and blue horns in scenarios (1) and (2) even in principle, 
since the line emission may be 
distributed axis-symmetrically.
Moreover, if $E_{\rm obs}<E_{0}$ then the observed line may
be {\it either} the red horn {\it or} blue horn of the full line profile
(but if $E_{\rm obs}>E_{0}$ the observed line is {\it always} the
blue horn since gravitational and transverse energy shifts can
take a blue horn below $E_{0}$ but can never take a red horn above
$E_{0}$).

In scenarios (1) and (2) above, the 
only way to avoid fine tuning is to have $r/r_{g}$  large enough
and $\theta$ small enough that the observed 
center energy of a particular line energy is never
very far away in energy from the rest value. 
However, in this part of parameter space,
scenarios (1) and (2) go smoothly to
scenario (3) anyway, in which the red and blue horns 
are so close in energy to each other that they are not resolved.
This also requires that the 6.607~keV and 6.931~keV  
lines must indeed be associated with
\fexxv and \fexxvi (see \fighorns).
Also, in the regime of large radius and small
inclination angle, scenarios (1) and (2)
require that the \fexxv and \fexxvi $K\alpha$ lines must
originate from the receding side of the
disk only in order to produce the observed redshifts. 

Thus, since scenarios (1) and (2) require such
highly contrived circumstances and fine-tuning, scenario (3) 
is the most likely
to be applicable to the He-like and H-like \fekalfa lines
in the NGC~7314 data. In that case,
no fine-tuning is required, and in particular, the
line emission does not have to be concentrated in a
narrow range of radii, nor is there any special
requirement about the azimuthal distribution of line
emission from the disk. 
In any case, our conclusion is testable with
a second \chandra observation: the center energies
of the 6.607~keV and 6.931~keV (low-state) lines should be
significantly different next time if scenarios (1) or (2) above are
applicable. This is because scenario (1), a hot spot 
origin, requires special
values of radius {\it and} azimuthal angle
of the hot spot on the disk, whilst scenario (2), emission
from an annulus, requires a special radius.
It is of course highly improbable that we would observe a
hot spot at the same radius, let alone at the same azimuthal
angle as well. 
Therefore, in a second observation, 
if there were no measurable difference in the line
energies relative to the first observation, and if scenario (2)
applies,
one would have to conclude that there is a special radius
and that we are again observing only the red Doppler
horn of a line profile originating in the same (receding) half
of the disk as in the first observation.
A far more plausible interpretation of immeasurable
differences in the line energies from observation to
observation would be that the red and blue Doppler
horns are not resolved, as in scenario (3) above.
Not too far in the future, \astroe will measure 
the narrow line profiles directly with a factor $\sim 6.5$ better
spectral resolution and settle the issue.
Until we obtain evidence to the contrary, we will assume
scenario (3) for the He-like and H-like \fekalfa lines in NGC~7314.

\subsubsection{Origin of the 5.842~keV Feature}
\label{lowefeature}

The above arguments against a hot-spot 
or annulus origin of the 6.607~keV and 6.931~keV
lines do not apply to the 5.842~keV feature
because it is not close to the rest energy of any \fekalfa line. 
However, it is easy to rule out a hot-spot origin
for the 5.842~keV feature. The most significant detection
of this line
is from the low-state spectrum, which is made of time
intervals covering a total time period of $\sim 100$~ks
(the length of the observation). Therefore, by definition,
a hot spot must have $r>>330r_{g}$, otherwise we would be
integrating over more than orbit and we would have an
annulus of emission, not a hot spot. However, in order
to get sufficient gravitational and Doppler redshifting,
\fighornse shows that 
we must have $r<132r_{g}$ (this upper limit is for
a rest energy of 6.400~keV; 6.966~keV gives an even smaller
upper limit).

As mentioned in \S\ref{geometry},
the difference in energy
between 6.400~keV and 5.842~keV is within 27~eV of the
difference between 6.931~keV and 6.400~keV.
Could the 5.842~keV and 6.931~keV lines correspond
to the red and blue horns, respectively, of an \fekalfa line
from an annulus? We can solve the equations for gravitational
and Doppler shifts (in a disk around a Schwarzschild black hole)
for values of $r/r_{g}$ and $\theta$ as a function
of the line rest energy, $E_{0}$. Solutions only exist for
values $E_{0}$ greater than 6.409~keV. For this rest energy,
$\theta = 90^{\circ}$ and $r=136r_{g}$. Since NGC~7314 is 
a type~1 AGN, smaller inclinations are more plausible and 
$E_{0}=6.966$~keV gives $\theta = 20.4^{\circ}$ and $r=16.4r_{g}$.
However, such an interpretation still requires fine tuning
since $\theta$ and $r$ have to be very special to give a
Doppler blue horn so close to the rest energy of \feklyap.
Moreover, since the ion fractions of \fexxv and \fexxvi
are so intimately related, one would expect to
observe a red horn corresponding to the 6.607~keV line;
the latter in this scenario would have to identified with the blue
horn of a He-like \fekalfa line. A red horn is predicted between
5.54--5.61~keV (depending on which He-like transition is applicable).
However, a narrow Gaussian emission line fitted at 10~eV intervals in the
5.0--6.0~keV range gives a reduction in the fit statistic, $\Delta C$,
which is never greater than 4.

On the other hand, the 5.842~keV line could still originate in an annulus
if the 6.931~keV line has nothing to do with the 5.842~keV line.
The constraints are shown in \fighorns (e) and \fighorns (f). These
figures show that for small enough radii (and correspondingly small
inclination angles), the 5.842~keV line could actually be the {\it blue}
Doppler horn of a wider line profile. The transition radius,
between the line being a red or blue Doppler horn, is $17.1r_{g}$
for $E_{0}=6.400$~keV and $9.3r_{g}$ for $E_{0}=6.966$~keV.
We cannot resolve the ambiguity in the parameters without more information.

\subsubsection{Comparison with NGC~3516}
\label{cmptjt}

Narrow lines attributed to localized \fekalfa emission from
an accretion disk were reported by Turner \etal (2002)
in another Seyfert~1 galaxy, NGC~3516, at energies
of $\sim 5.57\pm 0.02$, $6.22\pm 0.04$, $6.53\pm 0.04$, $6.84\pm 0.01$,
and $6.97\pm 0.06$~keV (in addition to the main peak at $6.41\pm 0.01$~keV).
The 6.84~keV and 6.97~keV lines are close to the rest energy
of \feklya so the same arguments given in \S\ref{narroworigin} with
respect to the 6.931~keV and 6.607~keV lines in NGC~7314
apply (namely, that their origin in hot spots or annuli is
unlikely). The remaining lines are likely to be due
to enhanced emission at particular radii, as discussed in Turner \etal (2002).

There is no conflict between the results and interpretation of
the narrow \fekalfa emission lines in NGC~7314 and NGC~3516.
Nor is there any conflict in the different interpretation of the
width of different lines in the same source. Line emission 
integrated over the surface of a disk which has
a flat radial line emissivity and which is viewed
at small inclination angles will naturally give rise to an
emission line profile which is narrow and which peaks close
to the rest energy of the line if the outer radius of emission
is large (several hundred $r_{g}$). Enhanced emission from
localized regions
on the same disk will give narrow emission lines which could
be hundreds of eV away from the line rest energy.

\subsection{Relation Between Energy Shift and Width of Disk Lines}
\label{widthshift}

We argued in \S\ref{narroworigin} above, that the
most likely reason for the \fexxv (f) and \feklya lines
in NGC~7314 to be narrow is that they originate
in an accretion disk which is nearly
face-on and which has a flat radial line emissivity. In that case
the red and blue Doppler horns are unresolved by the
\chandra gratings. 
For emission lines originating
in a disk rotating around a central mass, if the
Doppler horns are unresolved, there
is a definite relation between the minimum width, $W_{\rm min}$, 
of a line and its shift in centroid energy, $\Delta E$, for a given
disk inclination angle and {\it maximum} emission radius.
Note that the line emission may extend down to
small radii but the {\it minimum width} corresponds
to the separation of the extreme red and blue Doppler
horns at the maximum radius. {\it Independent of the radial emissivity
profile of the line}, the line width cannot be
smaller than this. Recall that we have rejected
a hot-spot or annulus origin of the lines.
We calculate $W_{\rm min}$ and $\Delta E$
in a very simple-minded way as follows.
We assume a Keplerian disk rotating around a Schwarzschild black hole,
and calculate, for a given disk inclination angle, and line emission from
a ring at a given $r/r_{g}$ ($r_{g} \equiv GM/c^{2}$), the smallest and
largest energy shifts measured by a distant observer (also taking
into account gravitational redshifting). 
The minimum width as a function of \rrg is then taken simply as the difference
between the two extreme  energies
(i.e. the separation of the red and blue Doppler horns), and the energy shift
is simply calculated as the average of these two extreme energies.
The true centroid energy should take account of the
differences in intensities of the red and blue horns
due to relativistic effects. However, we cannot
currently measure the shape of the lines in question so the approximation
is adequate for our purpose. 
Note that our assumption of a monochromatic line makes
$W_{\rm min}$ an even firmer lower limit on the width.
For example, the Fe~{\sc i}~$K\alpha$ and \feklya lines each consist of two lines,
separated by $\sim 13$~eV, and $\sim 19$~eV respectively
(the higher-energy component having twice the intensity of the lower-energy
one).

If $E_{0}$ is the rest energy of an emission line we
get, in the limit $r/r_{g} \gg 1$, $w \equiv (W_{\rm min}/E_{0}) 
\sim 2\sin{\theta} \sqrt{(r_{g}/r)}$ and $\epsilon \equiv (\Delta E/E_{0})
\sim (r_{g}/r)(1.5 - \sin^{2}{\theta})$, where $\theta$
is the disk inclination angle. From these equations,
one can eliminate $r$ to obtain a relation between the
minimum line width and the energy shift for a given
disk inclination angle.  
It is clear that if any lines in the Fe-K region
are unresolved by the \chandra HEG ($w <0.005$), there is already
a very severe constraint on the disk inclination angle
(it has to be less than a few degrees), and the maximum emission radius 
($r/r_{g} \gg 1$).
Now if $\sin{\theta} \ll 1$ 
(disk near face-on),
we get the simple result that $w \sim \epsilon^{\frac{1}{2}} \sqrt{(8/3)} \sin{\theta}$.  
Using this, and the low-state measurements of energy shift, and
width upper limit on \feklya (\tablehegfitsp), gives the most
conservative upper limit on the inclination angle.
The best-fitting $\epsilon$ gives $\theta < 5^{\circ}$ and
a maximum radius of $\sim 300r_{g}$. Using 
the lower limit on $\epsilon$ gives $\theta < 7^{\circ}$.
Using the latter upper limit on $\theta$ then gives an 
upper limit on the maximum radius of $\sim 610r_{g}$.

The exact relation
(i.e. not assuming $\sin{\theta} \ll 1$
or $r/r_{g} \gg 1$) between minimum line width and energy shift
is shown in \figwidthshift
for various inclination angles
($1000w$ is plotted against $1000 \epsilon$).
The radius goes from $10^{4} r_{g}$ to $6r_{g}$ 
for each curve, from left to right respectively. 
The width/shift curves in \figwidthshift have a very general
and simple interpretation for line emission from
a disk {\it integrated over arbitrary radii, regardless
of the radial emissivity of the line emission}. That is,
for a given disk, any integrated emission line
must lie {\it above} the curve corresponding
to the correct value of $\theta$ for that disk. The
measurements for a particular line plotted on the
diagram cannot lie below the curve corresponding to
the correct $\theta$
because large energy shifts in the line
centroid must be accompanied by correspondingly
large minimum line widths (i.e. separation of the
Doppler horns). Therefore if we observe a particular
emission line and its measurement errors place
it below a line of given $\theta$, then {\it all}
values of $\theta$ greater than that are ruled out.
Thus, the diagram is a powerful diagnostic.

The width upper limit and energy shift measurements 
for the \feklya line in the low-state spectrum of NGC~7314
are shown \figwidthshiftp.
It can be seen that \feklya alone immediately
rules out 
inclination angles $>  7^{\circ}$, consistent
with the simple estimate above.
The measured upper limit on the width of \feklya is very
firm (it is unresolved by \chandra), as is shift from its rest-energy, so
this line provides the tightest
limit on the inclination angle. 
We include \fexxv (f) in \figwidthshift
even though its identification is tentative.
As it happens, this line does not violate the
constraint $\theta < 7^{\circ}$, but it also does
not improve upon it.

\subsection{Timescales, Disk Ionization Structure, and Radial Line Emissivity}
\label{timescales}

We have shown that the
ionization balance of Fe in the accretion disk responds
rapidly to the continuum, on timescales $<12.5$~ks.
In the low state, the spectrum is
dominated by $K\alpha$ lines from
\fexxv, \fexxvip, and a line  at $\sim 6.4$~keV
from Fe~{\sc i}--Fe~{\sc xvii}.
As the continuum intensity increases the line emission
becomes complex, with more ionization stages between
Fe~{\sc i} and \fexxv contributing to the emission.
We cannot tell how much of the line emission at $\sim 6.4$~keV
comes from distant matter (see \S\ref{energyshifts}). 
However, if most of it comes from the disk, then a
multi-phase medium is implied since \fexxv and \fexxvi
cannot co-exist with significant ionization fractions of
Fe~{\sc i} or so. A two-phase medium is indeed predicted
by ionized disk models, which produce a warm/hot skin
above an underlying cooler region (e.g. Nayakshin and Kallman 2001). 
The relative thickness of the skin is dependent upon, among
other things, the shape of the X-ray continuum and the ratio
of X-ray to disk flux.
We note that the radiative recombination timescale for
\fexxv is $\sim 10$~($10^{10} \rm \ cm^{-3}/n_{e}$) 
($T/10^{6} \ \rm K$)$^{0.73}$ seconds (Shull \& van Steenberg 1982),
which is less than our 12.5~ks sampling time if
the electron density ($\rm n_{e}$) is not too small, or the temperature 
($T$) not too
high. In any case the temperature must be less than
$\sim 10^{9} \ K$ since the \feklya line is not resolved by 
the HEG. 

Thus, it appears that when the continuum luminosity 
increases, either cooler material (e.g. the underlying
disk) heats up, or warm/hot
material (e.g. the skin) cools down, in order to produce more
emission from the lower to intermediate ionization states
of Fe. To obtain the full picture one must observe
the X-ray continuum out to higher energies, especially
if the spectral shape varies with luminosity.
In fact, we show in \S\ref{pcafits} that the X-ray
spectrum is steeper in the high state than in the low state
so that what we call the high state may have a lower
ratio of hard X-ray to soft disk luminosity than the low
state, when the broadband spectrum is taken into consideration.

We note that
the hydrostatic timescale is $\sim 4.5 \alpha (r/20r_{g})^{\frac{3}{2}}$~ks
(e.g. see Nayakshin \& Kazanas 2002). At $6r_{g}$, this is $\sim 740$~s.
The X-ray continuum in NGC~7314 varies on a timescale of hundreds of seconds,
so the physical structure of most of the disk never
has time to adjust itself in response to the X-ray continuum
variability.
We showed in \S\ref{narroworigin} that the \fexxv~(f) and \feklya
lines in NGC~7314 likely originate in extended region on
the disk, rather than a hot spot or localized annulus.

We emphasize again that the \fexxv~(f) and \feklya
emission may extend down to a few gravitational radii
and still result in narrow emission lines (unresolved by
the \chandra HEG).
This is possible if the line emissivity (intensity per unit area)
emissivity falls off with radius less steeply than $r^{-2}$
(and may even be approximately uniform over some range in radius).
The X-ray continuum source is likely to be distributed over the whole
disk, possibly in the form of coronal flares (e.g.
Haardt \& Maraschi 1991, 1993; Svensson \& Zdziarski 1994; Merloni \& Fabian 2001).
The flat emissivity could result from, for example,
a roughly constant
number of coronal flares of roughly equal intensity per unit
area. A uniform illumination of the disk means that
the line profile will be dominated by emission from
large radii because of the larger area at larger radii.
Thus, the red wing of the line profile, expected from the
effects of strong gravity at the smallest
radii for steeper emissivity laws, will be very weak. A small disk inclination angle
then ensures that the entire line profile remains narrow.
This is illustrated in the
inset in \figwidthshiftp, which shows the line profile,
assuming monochromatic \feklya (in the disk rest frame), from a uniformly emitting
disk inclined at $3^{\circ}$, with the emission extending
from $6r_{g}$ to $400r_{g}$, calculated using the method of Fabian \etal (1989).
These parameters are a set consistent with the data.
The dotted lines show the
measured upper limits on the FWHM of \feklya (\tablehegfitsp).
It can be seen that the
red wing may be too small to detect even with higher spectral resolution.

Ionized disk models generally predict that the
He-like \fekalfa line should be much stronger than the \feklya line
(e.g. Ballantyne \etal 2001; Nayakshin \& Kallman 2001).
This is because the models do not include turbulence so
\feklya is suppressed by resonant scattering.
The fact that we observe the He-like and H-like lines
to have about the same EW ($\sim 60-70$~eV in the low state),
immediately tells us that turbulence must be significant.
Invoking a region of parameter space
of the disk in which the warm skin is negligible
does not solve the resonant scattering problem because the H-like line
is made in the skin itself. 

Finally, we can also argue that the Thomson depth ($\tau_{T}$) of the
disk corona is probably much less than unity, simply from that
fact that \feklya is unresolved. Compton scattering in a
corona with non-negligible optical depth would broaden the line.
If the line were so broad that we only detect the
unscattered component, the original line would have to
a factor $e^{\tau_{T}}$ stronger than the observed line.
The observed EW is already close to the maximum that
can produced by ionized disk models, even when resonant scattering
is suppressed. However,
collisional excitation of \feklya could produce
a larger (unscattered) EW. We note that 
a very large EW of \feklya can be produced by bulk-Comptonization
models (Bautista \& Titarchuk 1999), 
but these are optically-thin and the observed
line would be very broad, which is not the case in NGC~7314.

\section{RXTE SPECTRAL ANALYSIS}
\label{pcafits}

Here we present spectral analysis of the low-state and high-state
\rxte PCA spectra (combining PCU0 and PCU2). The extraction of
these spectra was described in \S\ref{data}. As explained in \S\ref{data},
we shall utilize the 3--15~keV data only.
\figrxtea shows the ratio of the high-state spectrum to the low-state
spectrum (both spectra were background-subtracted first).
The ratio of 3--15 keV count-rates in the high-state spectrum to the
low-state spectrum is 1.47 and this value is shown as a dotted, horizontal
line in \figrxteap. It is clear that the high-state spectrum is softer
than the low-state one. It does not appear to be a simple
change in slope of the continuum, however. There is a bump 
between $\sim 9-12$~keV. 

The spectral resolution of the PCA in the Fe-K band is
a factor $\sim 30$ worse than the \chandra HEG so we take the
best-fitting parameters of the Gaussian emission-line models
fitted to the HEG low-state and high-state spectra,
and use them for spectral fitting to the corresponding PCA 
spectra. All the line parameters, including the intensities, will
be frozen at the values given in \tablehegfitsp, after correcting
for the difference in HEG and PCA normalizations (see \S\ref{crosscal}). 
For the low-state spectrum, the model involves four Gaussian emission-line
components (\fekalfa (I), \fexxvp, \feklyap, and the 5.842~keV 
line). For the high-state spectrum, we include an additional line,
namely the redshifted \fekalfa (II) component (see \S\ref{lowstate},
and \tablehegfitsp).
\tablepcafits shows, not surprisingly, that these line models, when added to a 
simple power-law continuum with floating photon index ($\Gamma$) and
normalization, give a poor fit to both low-state and high-state PCA
spectra. 

Below, we compare the \rxte data to various models
of the X-ray reflection continuum from the accretion disk.
However, it should be remembered that the X-ray continuum
in NGC~7314 varies on a timescale of hundreds of seconds,
yet the hydrostatic timescale is $\sim 4.5 \alpha (r/20r_{g})^{\frac{3}{2}}$~ks
(e.g. see Nayakshin \& Kazanas 2002). At $6r_{g}$, this is 740~s.
Thus, most of the disk will not be in hydrostatic equilibrium, 
wheres the models assume equilibrium has been achieved.

\begin{table*}[tbh]
\caption{\rxte PCA Spectral Fitting Results}
\begin{tabular*}{\textwidth}{@{}l@{\extracolsep{\fill}}lrr}
\hline
& & & \\
Model $^{a}$ & Parameter & Low State & High State \\
& & & \\
\hline
& & & \\
Power Law & $\chi^{2}$/(degrees of freedom)     &  48.2/25 & 90.9/25 \\

{\sc pexrav}$^{b}$ & $\chi^{2}$/(degrees of freedom)   &  38.4/24 & 67.1/24 \\

Power Law \& Edge & $\chi^{2}$/(degrees of freedom)  & 41.9/23   &  35.2/23 \\

{\sc pexrav}$^{b}$ \& Edge & $\chi^{2}$/(degrees of freedom)  &
                        31.5/22     &            32.9/22 \\
& $\Gamma$ & $1.73^{+0.08}_{-0.07}$ & $1.59^{+0.12}_{-0.04}$ \\
& $R$ (Reflection Fraction) & $1.03^{+0.59}_{-0.47}$ & $0.02^{+0.60}_{-0.02}$ \\
& Edge Energy (keV) & $9.53^{+0.60}_{-0.63}$ & $7.97^{+0.36}_{-0.23}$ \\
& $\tau_{\rm threshold}$ & $0.088^{+0.059}_{-0.055}$ & $0.222^{+0.056}_{-0.063}$ \\
& & & \\
{\sc pexriv}$^{b}$ & $\chi^{2}$/(degrees of freedom)  & 39.3/22  &  26.2/22 \\

& $\Gamma$ & $1.73^{+0.07}_{-0.07}$ & $1.88^{+0.09}_{-0.07}$ \\

& $R$ (Reflection Fraction) & $0.71^{+0.44}_{-0.36}$
        & $1.23^{+0.62}_{-0.41}$ \\

& $T_{\rm disk}$ (Kelvin) &  ($10^{4}$--$10^{6}$) $^{d}$
        & $>1.2 \times 10^{5}$ \\

& $\xi$ ($\rm erg \ \rm cm \ s^{-1}$) & 0.02 ($<56$) & $27^{+62}_{-15}$ \\
& & & \\

{\sc xion}$^{c}$ & $\chi^{2}$/(degrees of freedom) & 21.6/17   &  15.5/17 \\

& $\Gamma$ & $1.72^{+0.06}_{-0.05}$ & $1.90^{+0.06}_{-0.08}$ \\

& $L{\rm x}/L_{\rm disk}$ & 0.02 ($<0.42$) & 0.05 ($<4.5$) \\

& $\dot{m}/ \dot{m}_{\rm Edd}$ & 0.49 ($>0.27$) & 0.07 ($<0.36$) \\
        &               & OR 0.02--0.15$^{e}$ & \\

& & & \\
\hline
\end{tabular*}
{\small Spectral fits were performed in the 3--15~keV band.
Details
of the fits in \S\ref{pcafits} (see also \figrxtep).
Statistical errors are quoted for 90\% confidence, 1 interesting
parameter ($\Delta \chi^{2}+2.706$). \\
$^{a}$
Gaussian emission lines were included (except for the
{\sc xion} model),
with parameter values frozen at the best-fitting low-state
and high-state \chandra HEG
values (\tablehegfitsp), corrected for HEG/PCA
cross-normalization difference (see \S\ref{crosscal}). \\
$^{b}$ {\sc pexrav} and {\sc pexriv} are `cold' and ionized disk-reflection models
respectively, due
to Magdziarz \& Zdziarski (1995). These models produce only continuum, no lines. \\
$^{c}$ {\sc xion}: ionized disk reflection model due to Nayakshin \& Kallman
(2001).
{\sc xion} does not allow disk inclination angles $<18^{\circ}$
and predicts \fexxv line emission to
be much stronger than that from \fexxvi (\figrxtep), but the
two lines are observed to have about the same EW
(\fighegfekp). Therefore we omitted the 5.5--7.5~keV
regions of the PCA spectra. \\
$^{d}$ No constraint on the disk
temperature was obtained in the given range. \\
$^{e}$
90\% confidence range of a second
solution for $\dot{m}/ \dot{m}_{\rm Edd}$ in the low state.
}
\end{table*}

\subsection{`Cold' Reflection}
\label{coldrefl}

To the above power law plus Gaussian-lines model, 
we added a continuum component due to Compton-reflection in
`cold', solar abundance, optically-thick matter (using the XSPEC model {\sc pexrav} --
see Magdziarz \& Zdziarski 1995). We fixed the inclination angle
at the smallest value allowed by the model ($18^{\circ}$),
guided by our findings from the \chandra spectra. The direct,
power-law model continuum is exponentially cut-off with a roll-over
fixed at 100~keV. Thus, there is only one extra free parameter
relative to the simple power-law plus Gaussian-lines model described above, and
that is the `amplitude' of the reflected continuum, $R$, relative to
that expected from a time-steady X-ray source illuminating a
reflecting medium which subtends a solid angle $2\pi$ at the source.
The model improves the fits but the
high-state fit is still very poor (see \tablepcafitsp). The problem is that the
high-state fit leaves residuals in the 6--7 keV band, which,
if modeled by additional Fe~line emission would imply at least
twice the intensity of total Fe~line emission than observed
by {\it Chandra}. Also, the high-state spectrum has an
edge-like feature at $\sim 8$~keV which is not modeled
by the cold reflection model. In fact if we add a simple edge
model (two additional free parameters: the threshold energy,
and optical depth at threshold), we obtain good fits for
both the low-state and high-state spectra. The best-fitting
parameters are given in \tablepcafitsp. It can be seen that 
the high-state spectrum does not actually require any reflection
continuum and could just as well be fitted with a power-law
and an absorption edge.

\subsection{Ionized Disk Reflection: {\sc pexriv}}
 
We attempted to fit a {\it cold} Compton-reflection model because
the \fekalfa (I) line (see \tablehegfitsp) detected by \chandra
could originate in an optically-thick obscuring torus (viewed 
near face-on). \tablepcafits shows that such a reflection component
could be present in the low-state spectrum, and that if
(as expected) it does not respond to rapid continuum variability,
it is at least
consistent with being present in the high-state spectrum.
However, we do not take these results literally,
and the complexity added by the absorption edges that seem to be
required rather suggests that we should attempt to fit models
of reflection from an ionized medium.
In that case the reflection continuum is much more complicated,
and in particular the Fe~K edge can be deeper and higher in 
energy compared to reflection in neutral matter (e.g.
see Ballantyne \etal 2001; Nayakshin, \& Kallman 2001).
Therefore we removed the edges from the above model and
replaced {\sc pexrav} with {\sc pexriv} (the latter is
an ionized reflection model due to Magdziarz \& Zdziarski, 1995).

The model {\sc pexriv} computes only continuum and no line emission so
we retained the Gaussian emission lines, their
parameters still fixed at the best-fitting \chandra values
(corrected for the HEG/PCA cross-normalization difference -- see \S\ref{crosscal}).
The results are shown in \tablepcafitsp. The best-fitting models
and ratios of data to model for the low-state and high-state spectra
are shown in \figrxteb.
Good fits are obtained, but the low-state data still have
a marked deficit between $\sim 10-11$ keV relative to the
model. \tablepcafits shows that $R$ is consistent with
being the same ($\sim 1$) for both the low-state and the high-state 
spectra, within the errors.
The {\sc pexriv} model has two additional free parameters
compared to {\sc pexrav}: the disk temperature ($T$), and
ionization parameter ($\xi$). The remaining parameters
are kept fixed at their values in the {\sc pexrav} model.
We found that the
disk temperature is essentially unconstrained in the
low state (no value in the allowed model range of $10^{4}$--$10^{6} \rm
\ K$ is preferred), and that the ionization parameter
is not bounded below (upper limit, $56 \rm \ erg \ cm \ s^{-1}$).
On the other hand, for the high-state spectrum the disk
temperature is bounded below (at $1.2 \times 10^{5} \rm \ K$)
and $\xi = 27^{+62}_{-15} \rm \ erg \ cm \ s^{-1}$. These
results can be interpreted in terms of a reflection
continuum being dominated by matter which is more strongly 
ionized in the high state than in the low state.

\subsection{Ionized Disk Reflection: {\sc xion}}

Next, we replaced the {\sc pexriv} model with the {\sc xion} model
in XSPEC (see Nayakshin \& Kallman 2001). {\sc xion} calculates
the reflected spectrum from an X-ray illuminated disk self-consistently,
balancing the hydrostatic and thermal structure of the disk.
The model does not allow $R$ to be varied (it is effectively
fixed at unity). The inclination angle and exponential cut-off
energy of the direct power law were kept fixed at the previous values.
The inner and outer disk radii were fixed at $6r_{g}$ and $600r_{g}$
respectively. The model allows three choices of geometry:
a central point-source illuminating the disk (`lamp-post' model), a central
sphere illuminating the disk, or coronal flares illuminating the
disk. We found that the choice makes negligible
difference to the best-fitting parameters and no difference to our
conclusions, and we quote results for the coronal flare case.
The model also includes relativistic smearing. 

The two parameters that we allowed to float in the {\sc xion} model were the
X-ray to disk luminosity, $L_{\rm x}/L_{\rm disk}$, and the
accretion rate, normalized to Eddington, $\dot{m}/\dot{m}_{\rm Edd}$
(see Nayakshin \& Kallman (2001) for details).
Finally, the model
calculates line emission in addition to the reflection continuum.
However, this is problematic because the model (in common with
other disk illumination models) predicts \fexxvi line emission to 
be weaker than \fexxv emission, which is not 
what is observed in our data for NGC~7314 (the He-like and H-like
lines have about the same EW). The reason is that turbulence
is not included, so \feklya is suppressed by resonant scattering.
Also, the model does not allow small enough inclination angles
to predict the detailed Fe~line emission that we observe
(the minimum inclination allowed is $18^{\circ}$).
Therefore we fitted the PCA spectra omitting the 5.5--7.5~keV
region. The main purpose of these fits is to investigate whether
we can obtain better fits to the continuum than
{\sc pexriv}, which contains less ionization physics. 

The results
are shown in \tablepcafits and \figrxtec (the latter showing
the best-fitting models and data/model ratios for the low-state
and high-state spectra). The fits are excellent and
we can see that the problematic feature in the data at 
$\sim 10$~keV is now fitted by an Fe recombination edge (which is
absent from the {\sc pexriv} model). On the other hand,
we also see that the line emission predicted by the model
is different to that observed, with the $K\alpha$ emission
from \fexxv being much stronger than that from \fexxvip.
The ratio $L_{\rm x}/L_{\rm disk}$ is not well-constrained
by the data although the upper limits suggest that the high-state spectrum
may have a value up to an order of magnitude higher than the low state.
For $\dot{m}/\dot{m}_{\rm Edd}$ there appear to be two
solutions for the low state: either this ratio is in the range
0.02--0.15 or it is $>0.27$. For the high state, $\dot{m}/\dot{m}_{\rm Edd}
<0.36$. Thus, better data and modeling are required. 
What we can say is that 
the photon index of the power-law continuum
is flatter in the low state than in the
high state and when one considers the broadband continuum, it may
be this spectral variability which is driving the
ionization state of the disk. What we call the high state
(defined as such by the luminosity in the narrow 0.8--7~keV
band) may in
fact correspond to a smaller $L_{X}/L_{\rm disk}$ since the
high-state spectrum is steeper than the low state.
 
\section{THE STUDY OF BLACK-HOLE ACCRETION-DISK PHYSICS
USING NARROW Fe~$K\alpha$ LINES}

\label{missions}

One of the major goals of \fekalfa line studies in AGN is
to ultimately map the space-time near the event horizon of
a black hole. It is anticipated that this will be a
key area for {\it Constellation-X} to break ground.
A `yardstick' with which we can gauge the precision with 
which we can map the space-time near the event horizon
is the precision with which we can measure the black-hole
angular momentum, or spin ($a$). 
The problem is that we do not currently understand the relation between
the X-ray continuum and relativistic \fekalfa line emission
in AGN well enough to 
construct realistic reverberation models 
(e.g. Ballantyne \& Ross 2002; Shih, Iwasawa, \& Fabian 2002). 
It is widely recognized that we first
need to understand the ionization structure of the accretion
disk and {\it its} relation to the X-ray continuum.
Thus, our \chandra results for NGC~7314 are important because
this source has an X-ray continuum which has a dynamic range
of rapid variability large enough to be useful.
NGC~7314 exhibits ionization stages of Fe from $\sim$~Fe~{\sc i} 
up to \fexxvip, and the ionization structure of Fe
responds to the rapid continuum variability. 
These properties have not been demonstrated
for any other AGN.
If we want to use reverberation techniques to probe
the space-time near the event horizon of a black hole we need to 
understand the ionization physics from an observational point-of-view,
in order to build a theoretical understanding.
To achieve this, we had better
study NGC~7314 further, and identify other suitable sources too.
In addition to pursuing reverberation mapping, there 
is a parallel approach we can pursue, described below, which does not rely on
any knowledge of the relation between the X-ray continuum
and the line emission.

\subsection{Alternative Technique for Measuring  Black Hole Spin}
\label{bhspin}

Spectra with a FWHM resolution of $\sim 6$~eV, obtainable in the
very near future with {\it Astro-E2}, will 
determine whether our interpretation of the 
\fexxv and \fexxvi narrow \fekalfa lines in NGC~7314 
(i.e. emission from a large area of a disk
having a flat radial line emissivity)  is correct or not,
by directly measuring the line profile. 
Once verified,
the direct observation of narrow \feklya line emission
from the accretion disk
in NGC~7314 will represent a critical step
toward the goal of making precision measurements of
key parameters of an accretion disk/black-hole system
using future, high-throughput, high spectral resolution
instrumentation because some of the current limitations of
Fe-K line studies (aside from the variability problem), may be overcome. 

Now, the precision with which we can measure
the physical parameters of the accretion disk/black-hole system
from an \fekalfa line profile
(principally, the inclination angle, $\theta$, 
the inner and outer radii of emission, and the black-hole spin)
depends directly on how well we can measure the energies of
the red and blue Doppler peaks at the outermost radii of
emission, and the minimum and maximum energies of the 
extrema of the entire profile (corresponding to the innermost
radii of emission). The Doppler peaks corresponding to the
outer radii become broader (as does the entire line profile) and more poorly defined as
$\theta$ increases, and as the radial line emissivity 
becomes more and more weighted toward smaller radii.
At some point these peaks become so broad that higher spectral
resolution does not help in improving the
precision with which they can be measured. 
When the lines are very broad (FWHM $\sim 1$~keV or so),
the maximum energy of the blue wing is the easiest to measure, 
but the minimum energy (the lowest energy of the red wing) is extremely
difficult to measure since it merges  smoothly with the continuum.
The extent of the red wing will always be ill-defined, 
no matter how good the spectral resolution is.
In addition to these problems,
if the \fekalfa line is due to the lower ionization states of Fe,
the uncertainty in the line rest energy will present an
inherent limit on the precision of measurement of the 
disk and black hole parameters. It should also be obvious that
\fekalfa lines from the lower ionization states of Fe are
subject to confusion with non-disk \fekalfa line emission from 
distant matter, which is difficult to disentangle without variability
information.

On the other hand, narrow \fekalfa lines from low-inclination disks
are affected significantly less by these problems. A flat
radial line emissivity is preferable (allowing larger inclination
angles to give narrow lines) but not essential. If we can identify
an emission line with \feklya (as we can in NGC~7314) then the
uncertainty in rest energy is eliminated. It will still be difficult
to measure the extent of the red wing, but in what follows we
will not require this to be measured.
All we have to do is measure the
energies of the two Doppler peaks corresponding to the outer radius
of emission ($r_{\rm max}$), 
and the maximum energy of the blue wing corresponding to
the approaching side of the disk at the inner radius ($r_{\rm min}$). 
All three
quantities will be clearly defined for a narrow line. Therefore
we have three measurements and four unknowns ($r_{\rm min}$, $r_{\rm max}$,
$\theta$, and $a$). Although the innermost stable orbit in a Kerr
metric is related to $a$, we cannot simply assume that the line emission
extends all the way to the last stable orbit. However, we
can say that $r_{\rm min}$ must be greater than the last stable radius.
The Doppler peaks corresponding to $r_{\rm max}$ have
a negligible dependence on $a$ if $r_{\rm max}>10r_{g}$. Thus,
we can solve for $r_{\rm max}$, and $\theta$, and
get an upper limit on $a$, or we can get a relation between
$r_{\rm min}$ and $a$.

Now, in addition to a narrow \fekalfa disk line, suppose we 
can identify a pair of red and blue Doppler peaks due to
enhanced emission at a localized radius
from the black hole. This radius must be less than
$\sim 10r_{g}$ in order to measure the spin, but 
whether this is the case or not
will be determined as part of the technique
(and the data rejected if this is not the case).
Then, measurement of the energies of the
red and blue Doppler peaks gives us two observables for 
only three unknowns (the emission radius, $r$, $a$, and the line rest energy), since
we have already measured $\theta$ from the main line profile,
as discussed above. If the original line was \feklya we may be able
to deduce its identity simply from the fact that for the given $\theta$,
the next lowest \fekalfa rest energy (at least 284~eV lower,
for the \fexxv resonance line) may not
be able to produce a blue peak at an energy as high as observed
whilst still satisfying the $r_{\rm min}$ versus $a$ relation
deduced from the main line profile. Then we can solve for $r$ and
$a$. If it is not possible to deduce the rest energy, we look
for another pair of red and blue Doppler peaks due to local
emission from a different radius. It could be at a completely
different time, even from a different observation, since
it is reasonable to expect that
$\theta$ and $a$ are not going to change. 
Then we will have
four observables for four unknowns (two emission radii, $a$, and
the rest energy) so we can solve uniquely for $a$. 
Obviously the line rest energy must be the same
for both events. If it is not the same we will know because we will not
be able to obtain a unique solution. In that case we look for further
events until we do obtain a unique solution.

As already mentioned, localized \fekalfa line emission has already
been observed in at least three AGN now (MCG~$-$6-30-15, Iwasawa \etal 1999;
NGC~3516, Turner \etal 2002; NGC~7314, 5.842~keV feature reported in this
paper). Although it has not always been possible to identify both
red and blue peaks of an event, the leap in effective area of
{\it Constellation-X} should show these features effortlessly.
In fact it would be more surprising if these localized events
were {\it not} observed. We expect them to be observed
with abundance by {\it Constellation-X}. It has been shown that, at least
for {\it Constellation-X}, even very modest localized enhancements,
of the order of only $\sim 10\%$ of the total line emission,
will be easily measurable in exposure times less than one day,
for the bright, well-known AGN (Yaqoob 2001). 
For particularly strong feature, we will be able to track the
center energy as a function of time and measure the black-hole
mass (e.g. Nayakshin \& Kazanas 2001).

It is important to realize that {\it nowhere in the above scenario
for measuring black-hole spin did we ever have to measure,
parameterize or know anything about the radial line emissivity},
even for measurements made from the main line profile.
All that is required is that the main line profile is {\it observed}
to be narrow,
and the central red and blue Doppler peaks have enough definition
to measure their energies with a precision matching the 
capabilities of the calorimeters on {\it Astro-E2} or {\it Constellation-X}.
The drop on the blue side of the line profile will automatically be
sharp under the above conditions. Although nothing about the
line radial emissivity needs to be known, if the line
profile is {\it observed} to have the above characteristics,
the implication is of course that the inclination angle is small and
the line radial emissivity flat.

\subsection{Observation of \feklya in Other Sources}
\label{felyaother}

It will now be necessary to re-examine and possibly re-interpret 
low spectral resolution data for AGN in general,
in the light of our findings. Certainly,
the iron line variability which was found by \asca to be very common
(e.g. Weaver,  Gelbord, \& Yaqoob 2001),
in particular the variable centroid energy, will be easier to interpret
in terms of variable proportions of emission-line intensities from different
ionization states. 
Other puzzling line variability results from \asca (e.g. 
Reynolds \& Nowak 2003 and references therein) need to be
re-examined. Ionized disk models make specific predictions
about the variability of the \fekalfa lines from 
different ions of Fe and direct comparison with the data
is much easier for the \fexxv and \fexxvi lines, especially
if they are not too broad (e.g. Ballantyne \& Ross 2002; Nayakshin \& Kazanas 2002).

There have been several reports of He-like or H-like
\fekalfa lines at low spectral resolution, using \asca,
\bsax, \xmm and even \rxte
(e.g. Nandra \etal 1996; Turner, George, \& Nandra 1998;
Comastri \etal 1998; Yaqoob \& Serlemitsos 2000;
Pounds \etal 2001; Reeves \etal 2001; Weaver \etal 2001;
Turner \etal 2001;
Matt \etal 2001; Reeves 2002; Perola \etal 2002).
Only in two of these cases has the \fekalfa line been identified
with \feklya (PG~$1116+215$, Nandra \etal 1996; Ton~S~180,
Turner \etal 1998, Comastri \etal 1998).
There is evidence for \feklya emission
in MCG~$-$6-30-15 from \xmm data (Fabian \etal 2002) but this
could not be distinguished from \fexxv resonance absorption.
These low-resolution data are subject to ambiguities from
line blending when the line emission is very broad.
For example, although optically-thin bulk-Comptonization models
have been invoked to explain the EW of the line in Ton~S~180
(Bautista \& Titarchuk 1999), blending of lines from other
ionization states of Fe cannot be ruled out. 

Fang \etal (2002) report an \feklya line
from the quasar H~$1821+643$ using the \chandra gratings, but is was
detected with a significance of only $2.3\sigma$. 
We examined \chandra \hetg data
for all type~1 AGN which are currently public and find marginal
evidence of \feklya in a few sources, but none as clear as NGC~7314
(Yaqoob \etal 2003, in preparation). 
It is worth remembering that the HEG effective area
in the Fe-K band is tiny ($\sim 15 \rm \ cm^{2}$ at
\feklyap). With future missions, which will have
larger collecting area and better spectral resolution
(such as {\it Astro-E2}, {\it Constellation-X}),
we should expect the \fexxv $K\alpha$ and \feklya emission lines to play an
important role in studying the black hole and accretion disk
in AGN. Even {\it Astro-E2} will be able to resolve all three
components of the He-like Fe triplet (see \S\ref{hetriplet})
and measure the \feklya to \feklyb
ratio (see \S\ref{felybeta}), thus improving considerably 
the constraints on the
ionization structure and physical conditions
in the accretion disk and corona.  

\section{SUMMARY}
\label{summary}

We observed the low-luminosity, narrow-line Seyfert~1 galaxy \src with the
\chandra \hetg (simultaneously with \rxte). The source exhibited
large-amplitude variability of the X-ray continuum,
by as much as a factor of $\sim 4$ in $\sim 3$~ks.
This is consistent with historical X-ray observations
and its relatively small central
mass of $5\alpha \times 10^{6} \ M_{\odot}$ (where $\alpha \sim 1$).
The most important aspect of our results is that
they establish NGC~7314 as a key laboratory for
future missions to study the ionization physics of accretion
disks. This is because the ionization structure of Fe varies rapidly
in response to the large-amplitude X-ray continuum variations,
and we detect ionization states of Fe from Fe~{\sc xvii} or lower, up to
\fexxvip. If we want to eventually map the space-time near the
event horizon of a black-hole using reverberation of 
\fekalfa lines, understanding the ionization physics of
the accretion disk is a pre-requisite. 

Below we summarize our results for NGC~7314.

1. We detected redshifted, unresolved emission lines from \fexxv (f),
\feklyap. Although the Fe He-like triplet is not resolved by
{\it Chandra}, if the emission is dominated by
the forbidden line, a consistent redshift is obtained for
both \fexxv and \fexxvi ($cz \sim 1500 \rm \ km \ s^{-1}$).
 
2. 
The least contrived geometry for the line-emitting region is that of a disk.
Other geometries are incompatible 
with the observational constraints and/or require much fine tuning.

3.
The most likely reason for the \fekalfa lines being narrow,
which requires the least fine tuning,
is that the disk inclination angle is small and the radial
line emissivity (intensity per unit area) is flatter than $r^{-2}$ (\S\ref{narroworigin}).
In that case the line width and redshift constraints measured
from \feklya alone imply that the line emission comes from within
$\sim 600$ gravitational radii
of the putative central black hole, and that the disk axis is inclined at $<7^{\circ}$
to the observer. 
Our interpretation of the data is easily testable
with further observations, even with current instrumentation.

4.
The \fekalfa line emission may extend all the way down
to a few gravitational radii of the black hole and yet still give an overall
line profile which is unresolved by {\it Chandra}, if
the radial line emissivity is flat.
A coronal X-ray source extending over the disk could
achieve this. This has important implications for AGN in general. 
If a disk is viewed at inclination
angles close to face-on, and the radial emissivity (intensity
per unit area) of an \fekalfa line falls off less steeply
than $r^{-2}$, the resulting emission-line profile
may not have a prominent red wing. The line may not even be broad,
even if the line emission
extends down to a few gravitational radii of the black hole.
Conversely, one cannot rule out a disk origin of an
Fe~$K\alpha$ line simply because it is narrow, even if it is
unresolved by the \chandra gratings.
The finding of a variable narrow \fekalfa line in Mkn~841
(Petrucci \etal 2002) supports these conclusions.

We emphasize that some AGN (such as MCG~$-$6-30-15) clearly
do have a strong red wing on the \fekalfa line and {\it do}
require a steep radial line emissivity. Our results
should be interpreted in the sense that we have now 
probably observed
the extreme ends of a distribution, with objects like
NGC~7314 at one end (Fe-K disk lines unresolved 
by {\it Chandra}), and objects like MCG~$-$6-30-15
(Fe-K disk lines several keV wide) at the other end.

5.
The fact that the \fexxv and \fexxvi $K\alpha$
lines have comparable equivalent widths implies that
turbulence is significant in the emitting medium, otherwise
\feklya would be suppressed due to resonant scattering.
Future missions will be able to reliably measure 
the \fexxvi Ly$\beta$:Ly$\alpha$ 
ratio and resolve the Fe He-like triplet,
and thus discriminate between different types of accretion models.

6.
The X-ray continuum varies on a timescale of hundreds
of seconds, which is faster than the hydrostatic
timescale even for the innermost regions of the 
accretion disk so most of the disk is never in hydrostatic
equilibrium. In the context of a disk-corona
model, the \feklya line must be formed in the ionized
disk, as the corona itself cannot produce the required equivalent width.
The corona is likely to be Thomson-thin, otherwise
the \feklya line would be broadened by Compton scattering.

7.
The ionization balance of Fe responds to continuum
variations on timescales less than 12.5~ks (which corresponds
to a light-travel distance of $\sim 500/\alpha$ gravitational
radii for $M = 5\alpha \times 10^{6} \ M_{\odot}$).

8.
Simultaneous \rxte data show that the continuum becomes steeper
as the continuum luminosity increases, so what we call the
high state (defined in terms of the 0.8--7~keV luminosity)
may in fact correspond to a smaller ratio of broadband X-ray to
disk luminosity, accounting for emission from lower ionization
states of Fe in the high state.

9. The \fekalfa line at $\sim 6.4$~keV
(unresolved, with equivalent width, $81 \pm 34$~eV) as a center energy
consistent with the $K\alpha$
line from Fe~{\sc i}--Fe~{\sc xvii}.
The offset velocity relative to the systemic velocity of NGC~7314
depends on which ionization stages dominate the line emission
(and is consistent with zero for Fe~{\sc i}).
It is not clear whether this component at $\sim 6.4$~keV originates
in the accretion disk, distant matter, or both, since there is
some evidence of rapid variability.
If all of this line originates in distant matter, the
upper limit on the width 
places it in the outer BLR or beyond. For a central mass of $5\alpha \times 10^{6}
M_{\odot}$ ($\alpha \sim 1$), this corresponds to
a lower limit on the distance of 
$>9700$ gravitational radii, or $>2.8\alpha$ light days from the X-ray source.

10.
When the X-ray luminosity in the
$\sim 3-10$~keV band is low, hardening of the X-ray continuum, above
$\sim 10$~keV, due to Compton reflection, is
greater than that in a high state.
The low state is dominated by reflection from low-ionization,
optically-thick matter, and the high state is dominated
by reflection from more highly ionized matter. The former
might be associated with distant-matter
and so would not respond to variability in the direct continuum.
The reflection from ionized matter is likely present all the
time, but responds to continuum variability and dominates the
spectrum in the high state. There is tentative evidence of
an \fexxvi recombination edge in the \rxte data.

11.
We presented details of a method by which future high throughput,
high spectral resolution instrumentation could measure
black-hole spin, without any knowledge or understanding
of the relation between X-ray continuum and \fekalfa line
emission (\S\ref{bhspin}). Using this method, the greatest
precision will be achieved when the line emission
from the accretion disk
is due to \feklya and when it is narrow.

The authors thank Peter Serlemitsos, Richard Mushotzky, Sergei Nayakshin, 
Martin Elvis, and
Barry McKernan for valuable discussions.
Support for this work was provided by NASA through \chandra Award Number
GO2-3133X issued by the Chandra X-ray Observatory Center, 
which is operated by the Smithsonian Astrophysical Observatory for and 
on behalf of the NASA under contract NAS8-39073.
The authors also gratefully acknowledge support from
NASA grants NCC-5447 (T.Y., U.P.), NAG5-10769 (T.Y.), and NAG5-7385 
(T.J.T). This research
made use of the HEASARC online data archive services, supported
by NASA/GSFC. This research has made use of the NASA/IPAC Extragalactic Database 
(NED) which is operated by the Jet Propulsion Laboratory, California Institute
of Technology, under contract with NASA.
The authors are grateful to the \chandra and \rxte
instrument and operations teams for making these observations
possible, and to an anonymous referee for helping to improve the paper.

\end{document}